\documentclass[preprint2]{aastex}

\shorttitle{HVF as the indicator of stellar population of SNe Ia}
\shortauthors{Meng, X.}

\begin{document}


\title{High-velocity feature as the indicator of the stellar population of Type Ia supernovae}


\author{Xiang-Cun Meng$^{\rm 1,2,3}$}
\affil{$^{\rm 1}$Yunnan Observatories, Chinese Academy of Sciences, 650216 Kunming, PR China\\
$^{\rm 2}$ Key Laboratory for the Structure and Evolution of
Celestial Objects, Chinese Academy of Sciences, 650216 Kunming, PR
China\\
$^{\rm 3}$Center for Astronomical Mega-Science, Chinese Academy of
Sciences, 20A Datun Road, Chaoyang District, Beijing, 100012, P.
R. China\\} \email{xiangcunmeng@ynao.ac.cn}





\begin{abstract}
Although type Ia supernovae (SNe Ia) are very useful in many
astrophysical fields, their exact nature is still unclear, e.g.
the progenitor and explosion models. The high-velocity features
(HVFs) in optical spectra of SNe Ia could provide some meaningful
information to constrain the nature of SNe Ia. Here, I show strong
evidence that the SNe Ia with strong CaII infrared triple (CaII
IR3) HVF around maximum brightness associate with relatively
younger population than those with weak CaII IR3 HVF, e.g. the SNe
Ia with strong CaII IR3 HVF tend to occur in late-type galaxy, or
early-type galaxy with significant star formation. In addition,
using pixel statistics I find that the SNe Ia with strong
maximum-light CaII IR3 HVF show a higher degree of association
with star formation index, e.g. ${\rm H}\alpha$ or near-UV
emission, than those with weak CaII IR3 HVF. Moreover, I find that
the strength of the CaII IR3 HVF is linearly dependent on the
difference of the absorption-weighted velocities between CaII IR3
and SiII 635.5 nm absorption lines, which then is a good index to
diagnose whether or not there is a high-velocity component in the
CaII IR3 absorption feature in the spectra of SNe Ia. I finally
discussed the origin of the HVFs and the constraints from our
discoveries on the progenitor model of SNe Ia.
\end{abstract}





\section{INTRODUCTION}\label{sect:1}
As the best distance indicator, the observation of Type Ia
supernovae (SNe Ia) leads to the discovery of an
accelerating-expansion Universe, which implies that there is a
mysterious dark energy in the Universe and the evolution of the
Universe is dominated by the dark energy(\citealt{RIE98};
\citealt{PER99}). Now, SNe Ia are becoming the center of a kind of
scientific industry, such as measuring the equation of dark energy
(\citealt{SULLIVAN11}; \citealt{MENGXC15}; \citealt{ABBOTT19}),
and constraining the cosmological model and basic physics via
comparing the Hubble constant value from SNe Ia with the one from
Planck satellite (\citealt{RIESS16}; \citealt{PLANCK18}). The
cosmological utility of SNe Ia as distance indicators is dependent
on one property of SNe Ia, i.e. their luminosity may be
standardized by the shape of their light curves
(\citealt{PHILLIPS93}; \citealt{RIESS96}; \citealt{PER97}).

Although SNe Ia are so important for cosmology and basic physics,
the exact nature of SNe Ia is still unclear. Generally speaking,
SNe Ia arise from the thermonuclear explosion of carbon-oxygen
white dwarfs (CO WDs) in binary systems (\citealt{HF60}), but we
still do not have an exact understanding on the explosion physics
and the evolution of the WDs towards explosion, i.e. various
explosion and progenitor models are proposed, but no any model may
explain all the properties of SNe Ia (\citealt{HN00};
\citealt{GOOBAR11}; \citealt{WANGB12}; \citealt{HILLEBRANDT13};
\citealt{MAOZ14}). Spectral features may provide clues for
diagnosing the explosion and progenitor models of SNe Ia
(\citealt{JHA19}), e.g. a variable or blueshift sodium line in
early spectra indicates that at least a part of SNe Ia may
originate in the single-degenerate systems (\citealt{PATAT07};
\citealt{STERNBERG11}), and a signature of the global asymmetry in
the innermost ejecta from later spectra implies an off-center
ignition in exploding WD (\citealt{MAEDA10}; \citealt{MAGUIRE18}).

Detailed spectroscopic analysis usually focuses on the absorption
features of silicon with minima which may indicate typical
photospheric velocities [the so-called `photospheric-velocity
feature' (PVF)], and be proposed as a diagnostic tool to inspect
the subcalsses of SNe Ia (\citealt{BENETTI05}; \citealt{BRANCH09};
\citealt{WANGXF09}). Another interesting feature in early SN Ia
spectra is the `high-velocity feature (HVF)', whose velocity is
higher than normal PVF by 6000 - 13000 ${\rm km~s^{\rm -1}}$
(\citealt{MAZZALI05a}; \citealt{MAGUIRE14}; \citealt{CHILDRESS14};
\citealt{ZHANGJJ16}). Since the first suggestion of HVF was made
by \citet{HATANO99}, it is found that almost all SNe Ia show HVF
in their early spectra (\citealt{MAZZALI05b}). The HVFs of SNe Ia
are often seen in CaII H\&K, Si II $\lambda$6355 and CaII infrared
triplet features (hereafter CaII IR3), and appear strong in their
early spectra, and become weak with time (\citealt{PARRENT12};
\citealt{MARION13}; \citealt{CHILDRESS14}; \citealt{ZHAOXL15}). In
addition, numerous spectropolarimetric observations show that the
line-forming region for the HVFs is physically distinct from that
for PVFs, and is substantially asymmetric
(\citealt{WANGLF06,WANGLF08}; \citealt{PATAT09};
\citealt{MAUND13}).

HVFs in SNe Ia have obtained an increasing attention and some
recent studies have found some correlations between different
observable quantities. For example, the strength of the HFVs was
discovered to correlate with the light-curve width of SNe Ia, i.e.
the SNe Ia with strong HVFs tend to have a wide light curve, while
the SNe Ia with narrowest light curve favor a weak HVF
(\citealt{MAGUIRE12,MAGUIRE14}; \citealt{CHILDRESS14}). However,
although there is a clue that CaII H\&K velocity correlates with
host galaxy stellar mass (\citealt{MAGUIRE12}), the strength of
HVFs from a larger sample seems not to be affected by the global
parameters of their host galaxies (\citealt{ZHAOXL15}).

Whatever, the physical origin of the HVFs and how the features are
correlated with the progenitor and explosion models of SNe Ia are
still lacking. Interaction with circumstellar material (CSM) was a
hot-discussing cause of the HVFs, where the CSM could be a clumpy
cloud, a torus or a shell (\citealt{KASEN03}; \citealt{WANGLF03};
\citealt{PATAT09}; \citealt{MULLIGAN17}). In addition, the HVFs in
SNe Ia may also be naturally explained by the explosion mechanism
itself, e.g. gravitational confined detonation on the surface of a
CO WD or the helium detonation in a CO WD envelope
(\citealt{PLEWA04}; \citealt{KASEN05}; \citealt{SHEN14}). No
matter what the origin of the HVFs, at least one of the abundance
enhancement, density enhancement or ionization effect at high
velocity region must be present (see the detailed discussion in
\citealt{MAZZALI05a})

However, although the sample of SNe Ia on the HVFs becomes larger
and larger, no determinative constraint on the progenitor and
explosion models of SNe Ia is obtained by studying the HVFs.
Especially, why wasn't the correlation between the strength of the
HVFs and the population of SNe Ia discovered since the strength
correlates with the width of the light curve (the width is an
index to represent the brightness of an SN Ia,
\citealt{PHILLIPS93}; \citealt{RIESS96}; \citealt{PER97}), while
it has long been established that bright SNe Ia favor young
stellar population and dim ones tend to belong to old stellar
population (\citealt{HAMUY96}; \citealt{WANGLF97};
\citealt{HOWELL01}; \citealt{SULLIVAN06}). In this paper, I will
show strong evidence that the SNe Ia with strong HVFs belong to
relatively young population, while those with weak HVFs favor
relatively old population.

In Sect.~\ref{sect:2}, I describe some definitions and the data
origin used in this paper. I will present my results in
Sect.~\ref{sect:3}, and discuss the origin of the HVFs and the
constraints of my discoveries on the progenitor models of SNe Ia
in Sect.~\ref{sect:4}. In Sect.~\ref{sect:5}, I will summarize
\textbf{my} main conclusions.

\begin{figure}
\centerline{\includegraphics[angle=270,scale=.38]{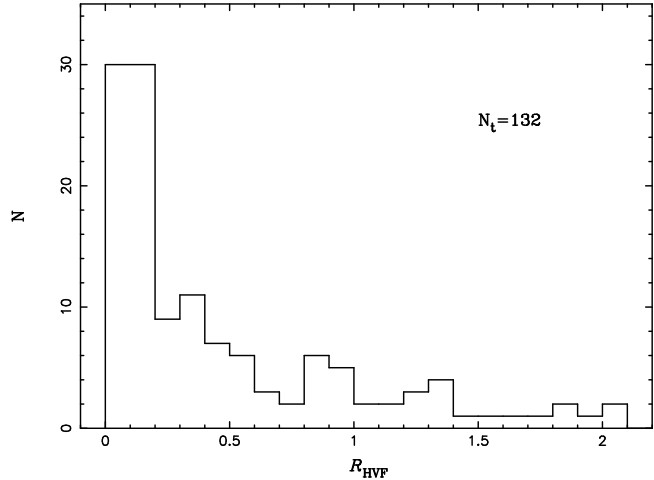}}
\caption{The distribution of the strength of the HVF from the CaII
IR3 absorption feature around maximum light (within 5 day of
B-band maximum brightness), where the data are from
\citet{SILVERMAN15}.}\label{rdis}
\end{figure}

\section{DATA}
\label{sect:2}
All the data used in this paper have been published in literatures
or may be easily obtained from NASA/IPAC Extragalactic Database
(NED). The majority of the SNe Ia used here are from Berkeley SN
Ia Program (BSNIP) and have been published in
\citet{SILVERMAN12b}. In this paper, I will mainly discuss whether
or not the strength of a HVF correlates with other observable
quantities. If there is a high-velocity component in an absorption
feature, the absorption feature may be fitted by two Gaussian
profiles, i.e. one is the HVF and the other is the PVF. Following
the definition in \citet{CHILDRESS14}, the strength of the HVF may
be described by the ratio of the pseudo-equivalent width (pEW) of
HVF absorption component to PVF absorption component, i.e.
 \begin{equation}
R_{\rm HVF}=\frac{\rm pEW(HVF)}{\rm pEW(PVF)}.\label{eq: rhvf}
  \end{equation}
Generally, the strength of the HVFs decreases with time, and
around the maximum light, the HVFs for some SNe Ia become very
weak, and even disappear, while the other SNe Ia still show very
strong HVFs. So, I choose the $R_{\rm HVF}$ value around the
maximum light as the indicator of the strength of HVF for one SN
Ia as did in \citet{CHILDRESS14}. CaII IR3 absorption line in the
spectra of an SN Ia usually shows the most remarkable HVF and is
widely studied. Therefore, I only focus on the CaII IR3 line, and
the $R_{\rm HVF}$ values of CaII IR3 lines for different SNe Ia
are from \citet{SILVERMAN15} and \citet{ZHAOXL15}.

\citet{MENGXC17} found that all SNe Ia may follow a universal
polarization sequence, i.e. the polarization of SiII 635.5 nm
absorption feature increases with a relative equivalent width
(REW), where the REW is defined as the ratio of the pEW to the
relative depth ($a$) of an absorption feature, i.e.
 \begin{equation}
{\rm REW}=\frac{\rm pEW}{a}.\label{eq: rew}
  \end{equation}
\citet{MENGXC17} suggested that the REW of SiII 635.5 nm
absorption line could be an indicator to diagnose the explosion
model of SNe Ia since the REW reflects the distribution of an
element in supernova ejecta. They found that the distribution of
the REW of SiII 635.5 nm around maximum light may be well fitted
by a Gaussian with an average value of 157.9 ${\rm \AA}$, and then
they suggested that their discovery could mean that all SNe Ia
share the same explosion mechanism, and only the delayed
detonation model has a potential ability to explain their
discovery at present. Here, following \citet{MENGXC17}, I will
check whether or not there is a correlation between the REW of
SiII 635.5 nm absorption line and the strength of the HVF of CaII
IR3 line around maximum light, where the values of pEW and $a$ for
different SNe Ia are mainly from \citet{SILVERMAN12}.

\citet{CHILDRESS14} defined an average absorption-weighted
velocity and found that the difference of absorption-weighted
velocity between CaII IR3 and SiII 635.5 nm lines correlates with
the decline rate [$\Delta m_{\rm 15}(B)$] of the light curve of
SNe Ia, where the absorption-weighted velocity is defined as
 \begin{equation}
\bar{v}=\frac{\int v\times a(v){\rm d}v}{\int a(v){\rm
d}v},\label{eq: va}
  \end{equation}
where $a(v)$ is the normalized absorption profile in velocity
space. \citet{CHILDRESS14} presented the absorption-weighted
velocities of the SiII 635.5 nm, CaII IR3, and CaII H\&K lines
around maximum light for 58 SNe Ia, and I will study whether or
not the absorption-weighted velocities correlates with other
properties of SNe Ia.

In this paper, I also want to check whether or not the strength of
the CaII IR3 HVF around the maximum light correlates with the
stellar population or stellar environment. The stellar environment
at the location of supernova explosion may reflect the information
of the stellar population of SNe Ia, and can be tested by checking
SN positions in their host galaxies and the global properties of
the host galaxies (\citealt{WANGXF13}; \citealt{ANDERSON15}). The
global parameters of the SN host galaxies, e.g. the physical size
and near-ultraviolet (nUV) absolute magnitude, are obtained from
NED\footnote{http://ned.ipac.caltech.edu/} and the position
information of SNe Ia in their host galaxies are from
\citet{WANGXF13} and \citet{ANDERSON15b}.

\citet{ZHAOXL15} found that the decay rate of $R_{\rm HVF}$
measured at 7 days before maximum brightness roughly correlates
with $\Delta m_{\rm 15}(B)$ and $V_{\rm max}^{\rm Si}$ (the
maximum-light photospheric velocity measured from SiII 635.5 nm
absorption line). Here, I also want to check whether or not the
decay rate of the $R_{\rm HVF}$ of CaII IR3 line around maximum
light correlates with the host galaxy morphology, the REW value of
SiII 635.5 nm absorption line and the strength of the CaII IR3 HVF
around maximum light, where the decay rate of the $R_{\rm HVF}$
around maximum light is calculated by

 \begin{equation}
\dot{R}_{\rm HVF}=\frac{\Delta R_{\rm HVF}}{\Delta t},\label{eq:
rdot}
  \end{equation}
where $\Delta R_{\rm HVF}$ is the difference of the $R_{\rm HVF}$
value of CaII IR3 line in two spectra between before and after
maximum ligh. Here, I only chose the SNe Ia whose CaII IR3 $R_{\rm
HVF}$ must have a value larger than 0 before maximum light, and
the spectrum phases must be within 5 days around maximum
brightness. The $R_{\rm HVF}$ values of CaII IR3 lines for
different SNe Ia are from \citet{SILVERMAN15}.



\begin{figure}
\centerline{\includegraphics[angle=270,scale=.38]{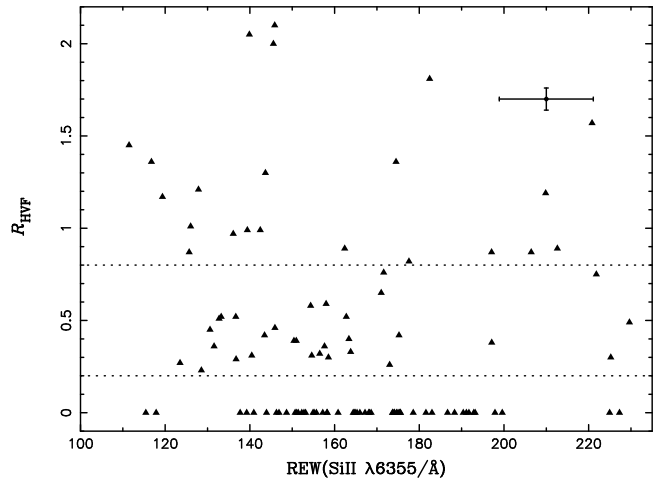}}
\caption{The strength of the HVF from the CaII IR3 absorption
feature vs the relative equivalent width of SiII 635.5 nm line
around maximum brightness, where the data are from
\citet{SILVERMAN12} and \citet{SILVERMAN15}. The two dotted lines
divide the sample into sub-samples with weak ($R_{\rm HVF}<0.2$),
strong ($R_{\rm HVF}>0.2$) and very strong CaII IR3 HVF ($R_{\rm
HVF}>0.8$). The black cross presents the typical
error.}\label{rhvf2}
\end{figure}

\section{RESULT}\label{sect:3}
\subsection{The distribution of $R_{\rm HVF}$ for CaII IR3 line}\label{sect:3.1}
\citet{SILVERMAN15} provided a very big sample of SNe Ia from
BSNIP, and 132 SNe Ia have a spectra with CaII IR3 absorption
feature around maximum light (within 5 day of B-band maximum
brightness). In their Table A4, the strength of the HVF of CaII
IR3 absorption line for the sample are presented, as defined in
Eq. (\ref{eq: rhvf}). In Fig.~\ref{rdis}, I show the distribution
of the strength of the CaII IR3 HVF ($R_{\rm HVF}$) around maximum
light\footnote{In \citet{SILVERMAN15} sample, the value of $R_{\rm
HVF}$ is set to be 0 if $R_{\rm HVF}<0.2$. So, I set the
distribution of $R_{\rm HVF}<0.2$ into one big bin in
Fig.~\ref{rdis}, in which 60 SNe Ia are included..}, where the
distribution shows a peak at low value and follows a long tail
until $R_{\rm HVF}>2.0$, i.e the stronger the CaII IR3 HVF, the
lower the frequency for an SN Ia to occur. Following the
definition in \citet{CHILDRESS14}, i.e. $R_{\rm HVF}>0.2$ means a
strong HVF, about 55\% of SNe Ia still present strong HVF in CaII
IR3 absorption feature around maximum brightness, and some even
show very strong HVF ($R_{\rm HVF}>0.8$). One question then
arises, i.e. what factor(s) lead(s) to the different strength of
the HVF between different SNe Ia. In the following sections, I
will present the possible answers for the question.

\begin{figure}
\centerline{\includegraphics[angle=270,scale=.38]{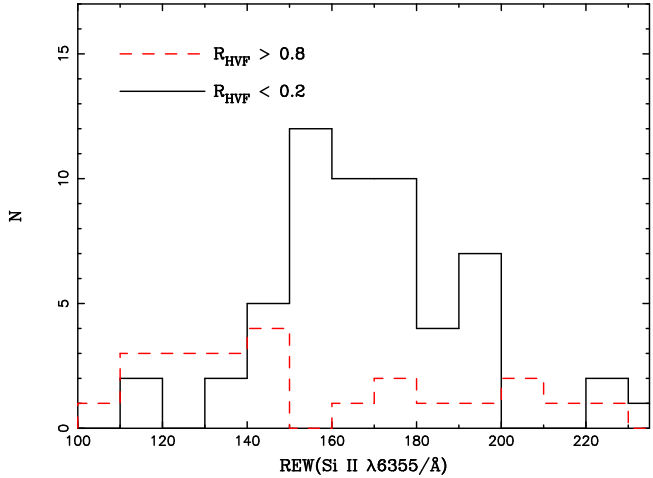}}
\caption{The number distributions of the relative equivalent width
of SiII 635.5 nm line for SNe Ia with weak (solid line) and very
strong CaII IR3 HVF (dashed line) around maximum light,
respectively.}\label{dishvf2}
\end{figure}

\subsection{The relation between $R_{\rm HVF}$ and REW}\label{sect:3.2}
The SiII 635.5 nm absorption line is the most remarkable feature
in the optical spectra of an SN Ia around the maximum brightness
and the REW of SiII 635.5 nm line could be an index reflecting the
key free parameters in the delayed-detonation model
(\citealt{MENGXC17}). Fig.~\ref{rhvf2} presents the relation
between the $R_{\rm HVF}$ of CaII IR3 absorption line and the REW
of SiII 635.5 nm absorption feature around maximum light. From the
first sight, there seems not to be a clear correlation between the
$R_{\rm HVF}$ and the REW, but the SNe Ia with weak maximum-light
CaII IRs HVF mainly distribute between REW=140 and 200, while
those with strong CaII IR3 HVF have a wide REW distribution.
Especially, the SNe Ia with very strong CaII IR3 HVF (i.e. $R_{\rm
HVF}>0.8$, \citealt{ZHAOXL15}) seem to disfavor the middle REW
value.

In Fig.~\ref{dishvf2}, I show the number distributions of the REW
for the SNe Ia with weak and very strong CaII IR3 HVF around
maximum light. The distributions are quite different, i.e. the SNe
Ia with weak CaII IR3 HVF focus on a value of the REW between 150
and 180, while those with very strong CaII IR3 HVF are relatively
rare in this region. A K-S test shows that the probability that
the two sub-samples are from the same mother sample is only
$1.1\times10^{\rm -3}$. \citet{MENGXC17} found that all kinds of
SNe Ia follows a universal polarization sequence, i.e. the
polarization of a SN Ia increases with the REW of SiII 635.5 nm
absorption line, and then they suggested that all SNe Ia could
share the same explosion model, no matter what their progenitors
are. If so, the different distributions in Fig.~\ref{dishvf2}
indicate that there would be another factor affecting the strength
of the HVFs in SNe Ia, e.g. different progenitors
(\citealt{WANGXF13}), rather than the explosion model itself.

If I check the Branch type of the SNe Ia (\citealt{BRANCH09}),
those with a weak CaII IR3 HVF around the maximum light favor the
Core normal (CN) and Cool (CL) SNe Ia, while those with a very
strong CaII IR3 HVF tend to be the shallow-silicon (SS) and
broad-line (BL) SNe Ia. Generally, 1991T-like SNe Ia belongs to SS
group, and 1991bg-like SNe Ia belongs to CL group
(\citealt{BRANCH09}). Actually, in the sample of
\citet{SILVERMAN15}, all of the eight 1991T-like SNe Ia have a
very strong CaII IR3 HVF around the maximum light (i.e. $R_{\rm
HVF}>0.8$), while among seventeen 1991bg-like SNe Ia, only 2002dk
shows $R_{\rm HVF}=0.38$ and the others present weak or no CaII
IR3 HVF around the maximum brightness. Such a discovery is
consistent with previous results, i.e. the SNe Ia with a fast
evolving light curve tend to show weak HVFs, and the SNe Ia with
strong HVFs usually have a slowly evolving light curve
(\citealt{MAGUIRE12,MAGUIRE14}; \citealt{CHILDRESS14}). It is well
established that 1991T-like SNe Ia arise from relatively young
stellar population and 1991bg-like SNe Ia belong to old stellar
population (\citealt{HOWELL01}; \citealt{JOHANSSON13};
\citealt{FISHER15}). This result implies that the strength of the
maximum-light CaII IR3 HVF could correlate with the stellar
population of SNe Ia. We will address this probability in the
following sections.

\begin{figure}
\centerline{\includegraphics[angle=270,scale=.38]{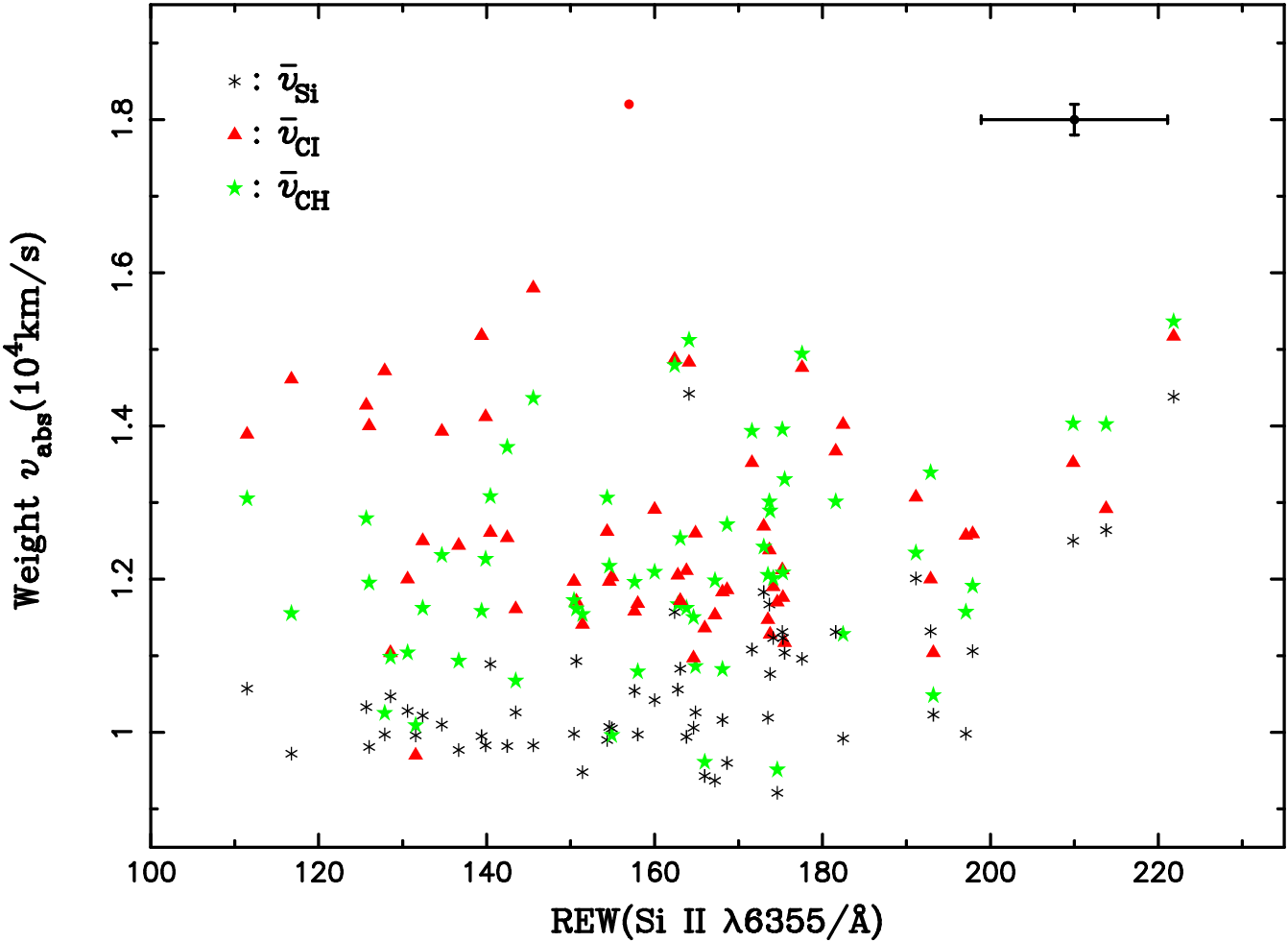}}
\caption{Absorption-weighted velocities for SiII 635.5nm, CaII IR3
and CaII H\&K features, respectively ($\bar{v}_{\rm Si}$, black
asterisks; $\bar{v}_{\rm CI}$, red triangles; $\bar{v}_{\rm CH}$,
green stars), versus the relative equivalent width of SiII 635.5
nm absorption line around the maximum light. The
absorption-weighted velocity data are from \citet{CHILDRESS14},
while the relative equivalent width data are from
\citet{SILVERMAN12}, and the cross represents the typical error of
the data.}\label{vhvf}
\end{figure}

\begin{figure}
\centerline{\includegraphics[angle=270,scale=.38]{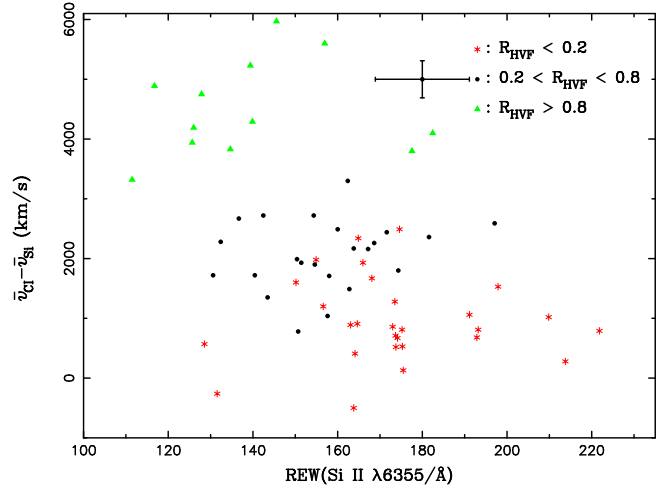}}
\caption{The difference between $\bar{v}_{\rm CI}$ and
$\bar{v}_{\rm Si}$ versus the relative equivalent width of SiII
635.5 nm line around the maximum brightness. Different points
represent the different range of CaII IR3 $R_{\rm HVF}$. The
absorption-weighted velocity data are from \citet{CHILDRESS14},
while the relative equivalent width data are from
\citet{SILVERMAN12}. The cross represents the typical error of the
data.}\label{vsi}
\end{figure}

\subsection{The relation between absorption-weighted velocities and REW}\label{sect:3.3}
\citet{CHILDRESS14} showed the dependence of the
absorption-weighted velocities for SiII 635.5nm ($\bar{v}_{\rm
Si}$), CaII IR3 ($\bar{v}_{\rm CI}$) and CaII H\&K ($\bar{v}_{\rm
CH}$) features and the difference between $\bar{v}_{\rm CI}$ and
$\bar{v}_{\rm Si}$ on $\Delta m_{\rm 15}(B)$ to verify that the
correlation between the strength of CaII IR3 HVF and $\Delta
m_{\rm 15}(B)$ are not artefact. In Figs.~\ref{vhvf} and
\ref{vsi}, I also show the dependence of these properties on the
REW of SiII 635.5 nm absorption line around the maximum
brightness. Fig.~\ref{vhvf} shows that $\bar{v}_{\rm CI}$ and
$\bar{v}_{\rm CH}$ do not significantly depend on REW, while
$\bar{v}_{\rm Si}$ seems to slightly increase with REW, i.e. no
significant physical correlation exists between the
absorption-weighted velocities and the REW of SiII 635.5 nm line.
However, Fig.~\ref{vsi} presents a clear trend that the larger the
difference between $\bar{v}_{\rm CI}$ and $\bar{v}_{\rm Si}$, the
smaller the REW of SiII 635.5 nm absorption line. Especially, the
SNe Ia with smaller REW (or larger difference between
$\bar{v}_{\rm CI}$ and $\bar{v}_{\rm Si}$) tend to have a stronger
HVF of CaII IR3 line as indicated in Fig.~\ref{rhvf2}. In
addition, Fig.~\ref{vsi} already indicates a fact that the
strength of the CaII IR3 HVF would depend on the difference
between $\bar{v}_{\rm CI}$ and $\bar{v}_{\rm Si}$.

\begin{figure}
\centerline{\includegraphics[angle=270,scale=.38]{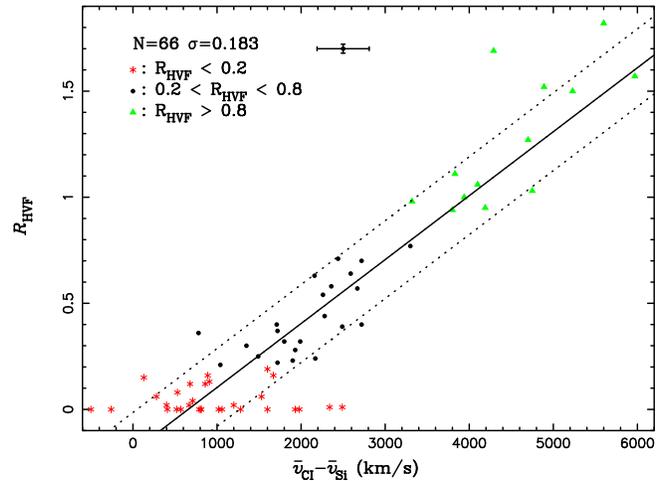}}
\caption{The relation between the strength of the HVF of CaII IR3
line and the difference between $\bar{v}_{\rm CI}$ and
$\bar{v}_{\rm Si}$ around maximum brightness. Different points
represent the different range of CaII IR3 $R_{\rm HVF}$. The solid
line shows a linear fit of the data and the two dotted lines
present the 1 $\sigma$ statistic error of the fit. The data are
from \citet{CHILDRESS14}, and the cross represents the typical
error of the data.}\label{rvsi}
\end{figure}

\begin{figure}
\centerline{\includegraphics[angle=270,scale=.38]{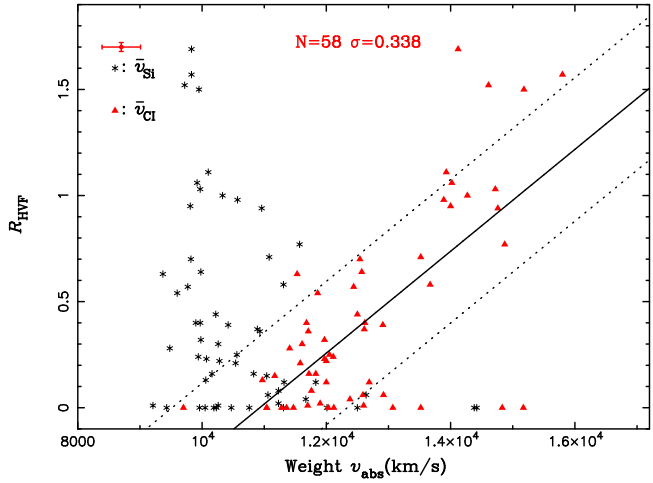}}
\caption{The relation between the strength of the HVF of CaII IR3
line and the absorption-weighted velocity $\bar{v}_{\rm CI}$ and
$\bar{v}_{\rm Si}$. The solid line shows a linear fit between
$R_{\rm HVF}$ and $\bar{v}_{\rm CI}$, and the two dotted lines
present the 1$\sigma$ statistic error of the fit. The data are
from \citet{CHILDRESS14}, and the cross represents the typical
error of the data.}\label{rva}
\end{figure}

In Fig.~\ref{rvsi}, I show the correlation between the strength of
the CaII IR3 HVF and the difference between $\bar{v}_{\rm CI}$ and
$\bar{v}_{\rm Si}$ around maximum brightness. There is a very good
linear relation between the strength of the HVF and the velocity
difference, i.e. the larger the velocity difference, the stronger
the CaII IR3 HVF [$R_{\rm HVF}=-0.198+3.013\times10^{\rm
-4}(\bar{v}_{\rm CI}-\bar{v}_{\rm Si})$]. Such a relation is
mainly from a fact that all SNe Ia have a similar
absorption-weighted velocity of SiII 635.5 nm absorption line
around maximum brightness, and then the strength of the CaII IR3
HVF is mainly dominated by its absorption-weighted velocity, as
shown in Fig.~\ref{rva}, which presents the relation between the
$R_{\rm HVF}$ of the CaII IR3 line and the absorption-weighted
velocity $\bar{v}_{\rm CI}$ and $\bar{v}_{\rm Si}$. Fig.~\ref{rva}
also shows a linear relation between the strength of the CaII IR3
HVF and $\bar{v}_{\rm CI}$, but the scatter of the linear fit is
much larger than that in Fig.~\ref{rvsi}, i.e. the difference
between $\bar{v}_{\rm CI}$ and $\bar{v}_{\rm Si}$ is the indicator
to measure the strength of the CaII IR3 HVF.

In addition, Fig.~\ref{rvsi} shows that some SNe Ia with $R_{\rm
HVF}=0$ have a large value of ($\bar{v}_{\rm CI}$-$\bar{v}_{\rm
Si}$), even larger than 2000 ${\rm km~s^{\rm -1}}$. Based on the
linear relation found Fig.~\ref{rvsi}, these SNe Ia would show
$R_{\rm HVF}\sim0.2-0.5$. Similarly, some SNe Ia without CaII HVF
also significantly deviate from the linear fit in Fig.~\ref{rva}.
Maybe, these SNe Ia also have a line-forming region for HVFs in
the supernova ejecta, but the region overlaps with or does not
distinct from the one for PVFs, which could be the reason why
these SNe Ia show a significantly larger photospheric pEW value
than other SNe Ia with weak CaII IR3 HVF, e.g. SN 2000dk, 2006X,
2006gt, 2007ba and 2007fr (see Table 2 in \citealt{CHILDRESS14}).
A piece of evidence supporting such an idea is from SN 2006X. In
\citet{CHILDRESS14}, the pEW of the photospheric CaII IR3 line in
the maximum-light spectrum of SN 2006X is $320\pm1{\rm \AA}$,
without HVF, but in \citet{SILVERMAN15}, the pEW of the
photospheric CaII IR3 line in the same spectrum is
$166.6\pm5.3{\rm \AA}$, with very strong HVF, i.e. $R_{\rm
HVF}=0.89\pm0.03$ (see also \citealt{ZHAOXL15}). Such a different
result could be derived from a fact that \citet{SILVERMAN15} have
a series of spectra on SN 2006X from a very early phase to a phase
after maximum light, while \citet{CHILDRESS14} only analyze one
spectrum at $t=2$ day. Based on the evolution of the spectra, it
would be relatively easy for \citet{SILVERMAN15} to judge whether
or not there is a high-velocity component in a CaII IR3 absorption
feature, and to estimate the possible strength of the HVF. So,
based on the result in Fig.~\ref{rvsi}, the difference between
$\bar{v}_{\rm CI}$ and $\bar{v}_{\rm Si}$ becomes a very good
indicator which may be helpful to judge whether or not there is a
CaII IR3 HVF in the spectrum of an SN Ia, i.e. if the difference
is larger than 1000 ${\rm km~s^{\rm -1}}$, it is likely that there
is high-velocity component in the CaII IR3 absorption feature,
even if a single Gaussian profile may fit the feature.

\begin{figure}
\centerline{\includegraphics[angle=270,scale=.38]{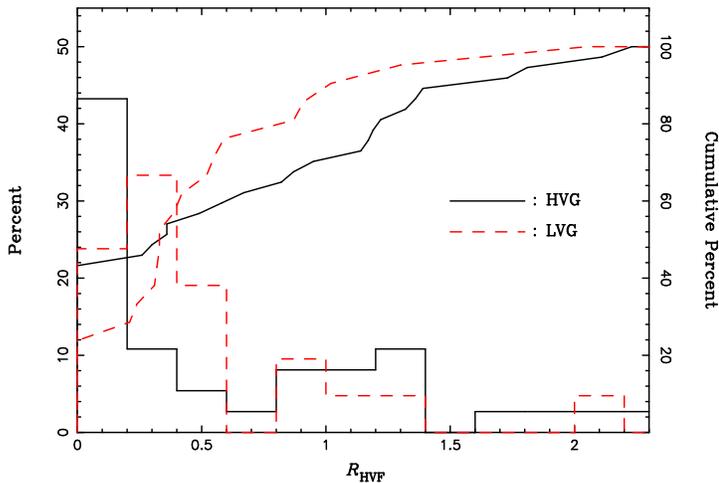}}
\caption{The number and cumulative precent distributions of the
$R_{\rm HVF}$ of CaII IR3 line around maximum brightness for HVG
(solid line) and LVG (dashed line) SNe Ia, respectively. The data
are from \citet{SILVERMAN15}.}\label{disvg}
\end{figure}

\subsection{The distribution of $R_{\rm HVF}$ for HVG and LVG SNe Ia}\label{sect:3.4}
Observationally, the photospheric velocity decreases with time,
but with different temporal velocity gradient. Based on the
different velocity gradient, \citet{BENETTI05} divided normal SNe
Ia into two sub-groups, i.e. high-velocity gradient (HVG) and
low-velocity gradient (LVG) SNe Ia. The different velocity
gradient could reflect the view angle relative to an asymmetric
explosion center, i.e. the LVG SNe Ia are viewed from the
direction of the off-center initial sparks, while the HVG ones are
viewed from the opposite direction (\citealt{MAEDA10}). In
Fig.~\ref{disvg}, I show the distributions of the $R_{\rm HVF}$ of
CaII IR3 line around the maximum brightness for HVG and LVG SNe
Ia, respectively. The two distributions look similar to that shown
in Fig.~\ref{rdis}, i.e. a peak at low $R_{\rm HVF}$ and a long
tail of high $R_{\rm HVF}$. Similarly, the difference of the
cumulative percent distributions of the $R_{\rm HVF}$ between HVG
and LVG SNe Ia are also not significant. A K-S test shows that the
probability that the two sub-samples are from the same mother
sample is as high as about 60\%. Then, the strength of the HVF of
CaII IR3 line around maximum brightness could not depend on the
view angle to the asymmetric explosion center, if the velocity
gradient reflects the view angle. This result could indicate that
the HVFs in the spectra of SNe Ia could not be derived from the
explosion mechanism of SNe Ia.

\begin{figure}
\centerline{\includegraphics[angle=270,scale=.38]{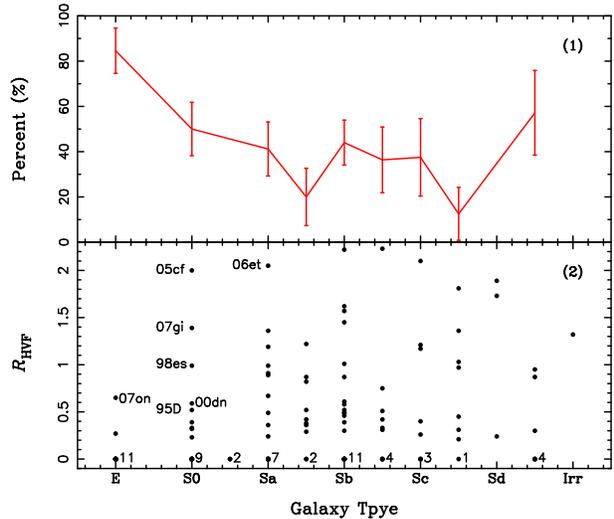}}
\caption{Panel (1) shows the number fraction of the SNe Ia with
$R_{\rm HVF}=0$ as a function of host galaxy morphology, and the
bars show the statistic error of the number fraction, assuming a
binomial distribution (\citealt{CAMERON11}). Panel (2) shows the
$R_{\rm HVF}$ of CaII IR3 line around maximum brightness versus
the galaxy morphology, where the spiral galaxies with and without
a bar are not discriminated. The numbers at the bottom represent
the number of SNe Ia with $R_{\rm HVF}=0$ in every host galaxy
morphology. The $R_{\rm HVF}$ data are from \citet{SILVERMAN15},
and the host galaxy morphologies are from \citet{SILVERMAN12b} or
NED.}\label{rtype2}
\end{figure}

\subsection{The dependence of $R_{\rm HVF}$ on the host galaxy type}\label{sect:3.5}
\citet{ZHAOXL15} checked the potential dependence of the number
distributions of SNe Ia with strong and weak HVFs on their host
galaxy morphologies, the K-band absolute magnitude of the host
galaxies and the normalized radial distance of the SNe Ia in their
host galaxies, and did not find significantly dependence on these
parameters. Such a result seems to suggest that there is not a
correlation between the strength of HVFs and the stellar
population (but see \citealt{PANYC15}). However, the number
distribution could not completely reflect the intrinsic dependence
of the strength on the stellar population. Here, I want to check
the potential dependence again by another way, i.e. directly
studying the relation between the strength of HVFs and the
parameters indicating the stellar population of SNe Ia. In
Fig.~\ref{rtype2}, I show the $R_{\rm HVF}$ of SNe Ia and the
number fraction of the SNe Ia with weak maximum-light CaII IR3 HVF
\textbf{($R_{\rm HVF}=0$)} as a function of host galaxy
morphology. In the host galaxy with a morphology later than S0,
the number fractions of the SNe Ia with weak maximum-light CaII
IR3 HVF are not significantly different with each other within
error, which is a similar result to that found in
\citet{ZHAOXL15}. However, the number fraction of the SNe Ia with
weak maximum-light CaII IR3 HVF in elliptical galaxies seems to be
higher than that in later type galaxies, which imply that the SNe
Ia with weak CaII IR3 HVF favor old stellar population. Most of
the SNe Ia with strong maximum-light CaII IR3 HVF tend to occur in
later type galaxies, in which the star formation rates are
generally high. Although a high star formation generally favor
core-collapse SNe, arising from short-lived progenitors, it may
also increase the birth rate of SNe Ia relative to the genuinely
old stellar population (\citealt{DELLA94};
\citealt{NAVASARDYAN01}). Then, the result here could indeed imply
that the SNe Ia with strong HVFs belong to relatively young
population.

Whatever, there are still some SNe Ia presenting a strong CaII IR3
HVF around maximum brightness in elliptical and lenticular
galaxies, e.g. SN 1995D, 1998es, 2000dn, 2005cf, 2007gi and
2007on. Generally, elliptical and lenticular galaxies are
passively evolving and don't contain any young stars. However, we
should keep in mind that many early-type galaxies present recent
star formations (\citealt{SALIM05}; \citealt{SCHAWINSKI07}). Then,
I checked the detailed circumstance of these SNe Ia. SN 2005cf has
a highest maximum-light CaII IR3 $R_{\rm HVF}$ among SNe Ia hosted
in lenticular galaxies and its host galaxy is a peculiar
lenticular galaxy (MCG -01-39-003), interacting with its neighbor.
There is a tidal bridge between MCG -01-39-003 and its neighbor,
and SN 2005cf locates close to the tidal bridge
(\citealt{PASTROELLO07}). It is widely known that the interaction
between galaxies may enhance the star formation rate. In addition,
the blue ultraviolet color of the galaxy from \emph{Galaxy
Evolution Explorer} (GALEX) also indicates a high star formation
rate in MCG -01-39-003 (\citealt{SMITH10}). Then, SN 2005cf would
associate with a relatively young population.

SN 2007gi is another one with very high maximum-light CaII IR3
$R_{\rm HVF}$ in a lenticular galaxy. Strictly speaking, the host
galaxy of SN 2007gi (NGC 4036) is not a typical lenticular galaxy,
but a morphology between S0 and Sa, with strong activity which is
another index to trigger the star formation in a galaxy
(\citealt{VERON06}; \citealt{ANN15}). Interestingly, SN 2007gi
just locates close to the inner activity region
(\citealt{ZHAOXL15}). For SN 1998es, although the morphology of
its host galaxy (NGC 0632) is classified into S0, the star
formation phenomena in the galaxy are very strong and it is a
star-burst galaxy (\citealt{BALZANO83}). The case for SN 2007on is
also very interesting, i.e. its host galaxy (NGC 1404) interacted
with its neighbor galaxy (NGC 1399) about 1.2 Gyr ago and there
are a large amount of intracluster medium between these two galaxy
(\citealt{SHEARDOWN18}). Now, NGC 1404 is falling into the center
of Fornax cluster and interacting with the intracluster medium to
form a sharp leading edge, where SN 2007on just locates close to
(\citealt{SUYY17}; \citealt{GALL18}). For the host galaxies of SN
1995D (NGC 2962) and 2000dn (IC 1468), there are rings/arms around
their main bodies and the rings/arms are significantly bluer than
their main bodies, which indicate the star formation activities or
relatively young stellar population in the galaxies
(\citealt{MARINO11}). These two NSe Ia locate either at a arm tail
or close to one rings (\citealt{ZHAOXL15}, see also the image in
SIMBAD\footnote{http://simbad.u-strasbg.fr/simbad/sim-fbasic}).
According to above detailed check, we find that these SNe Ia with
strong maximum-light CaII IR3 HVF in early type galaxies also
correlate with star formation activity or relatively young stellar
population (see also \citealt{PANYC15}). If these SNe Ia
correlating with young stellar population are eliminated, the
number fraction of the SNe Ia with weak maximum-light CaII IR3 HVF
in lenticular galaxies would also significantly higher than that
in later type galaxies, as shown in elliptical galaxies, i.e. the
SNe Ia with weak maximum-light CaII IR3 HVF favor old stellar
population.

\begin{figure}
\centerline{\includegraphics[angle=270,scale=.38]{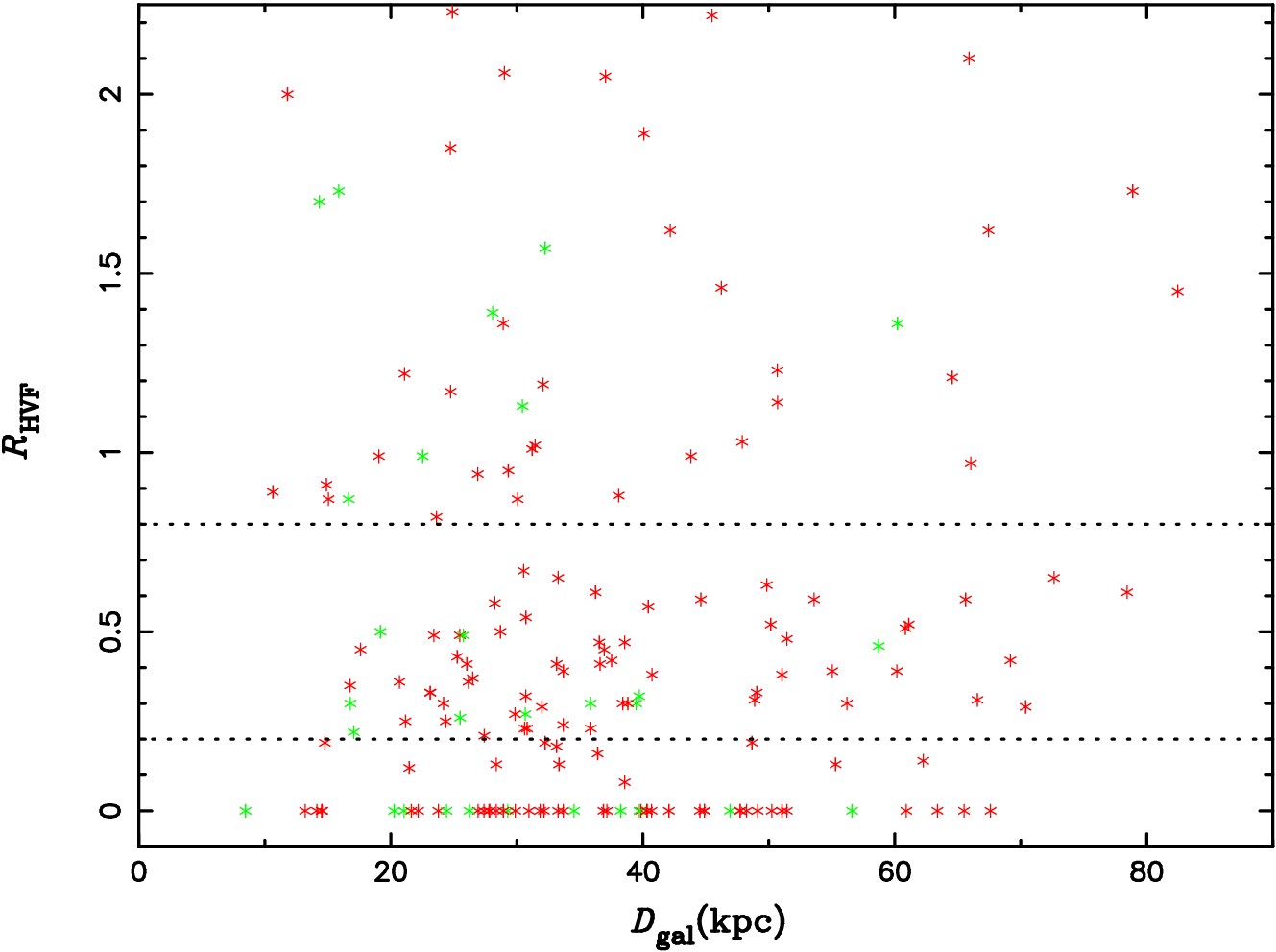}}
\caption{The CaII IR3 $R_{\rm HVF}$ around maximum brightness
versus the physical sizes (major axis) of host galaxies, where the
red points means that the sizes are represented by the B-band
light major axis at 25 mag ${\rm arcsec}^{\rm -2}$ isophote, while
the green ones by SDSS r-band major axis at 25 mag ${\rm
arcsec}^{\rm -2}$ isophote. The $R_{\rm HVF}$ data are from
\citet{SILVERMAN15} and \citet{ZHAOXL15}, and the physical sizes
of host galaxies are from NED. The two dotted lines divide SNe Ia
into weak ($R_{\rm HVF}<0.2$), strong ($R_{\rm HVF}>0.2$) and very
strong ($R_{\rm HVF}>0.8$) subgroups.}\label{rrad2}
\end{figure}

\begin{figure}
\centerline{\includegraphics[angle=270,scale=.38]{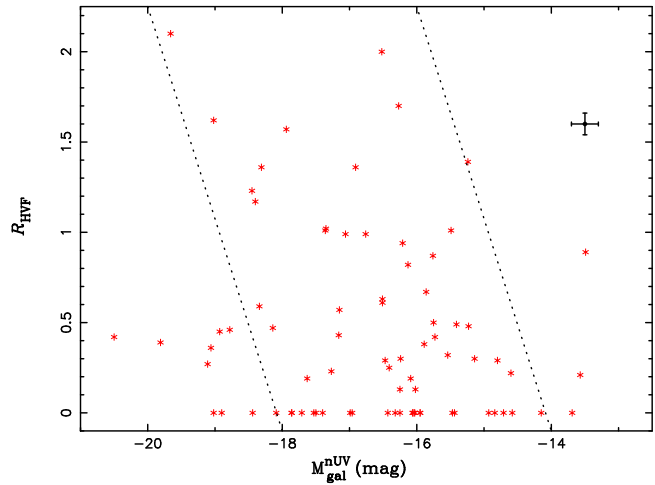}}
\caption{The CaII IR3 $R_{\rm HVF}$ around maximum brightness
versus the GALEX nUV band absolute magnitudes of host galaxies.
The cross shows the typical error of the data. The $R_{\rm HVF}$
data are from \citet{SILVERMAN15} and \citet{ZHAOXL15}, and the
nUV-band absolute magnitudes of host galaxies are from
NED.}\label{rmu}
\end{figure}

\subsection{The dependence of $R_{\rm HVF}$ on the global parameters of host galaxies}\label{sect:3.6}
\citet{ZHAOXL15} checked the potential dependence of the number
distribution of SNe Ia with different $R_{\rm HVF}$ on the K-band
absolute magnitudes of host galaxy, and no significant difference
between the strong and weak HVF samples is found. Here, I also
want to check whether or not the $R_{\rm HVF}$ value depends on
the other global parameters of the host galaxies, e.g. physical
sizes and nUV-band absolute magnitude. In Fig.~\ref{rrad2}, I show
the CaII IR3 $R_{\rm HVF}$ value of SNe Ia around maximum
brightness as a function of the physical size (major axis) of
their host galaxies, but I do not find any potential correlation
between $R_{\rm HVF}$ and the major axis, i.e. the number
distributions of the physical size of the host galaxies for the
SNe Ia with weak and strong maximum-light CaII IR3 HVFs are
indistinguishable.

In section~\ref{sect:3.5}, I find a clue that the CaII IR3 $R_{\rm
HVF}$ value of SNe Ia around maximum brightness could correlate
with their stellar population, but the number distribution of SNe
Ia with different $R_{\rm HVF}$ does not rely on the K-band
absolute magnitudes of their host galaxies (\citealt{ZHAOXL15}).
Such an inconsistency might be derived from that the K-band
absolute magnitude of a host galaxy is not a good index reflecting
the star formation in the galaxy, and the K-band light in a galaxy
is generally dominated by old stellar population
(\citealt{MANNUCCI05}). However, the ultraviolet absolute
magnitude of a host galaxy would be a better index in the galaxy
(\citealt{SALIM05}; \citealt{SCHAWINSKI07}). I obtained the GALEX
nUV absolute magnitudes of the host galaxies of SNe Ia in the
sample of \citet{ZHAOXL15} and \citet{SILVERMAN15} from NED to
check the potential correlation between the maximum-light CaII IR3
$R_{\rm HVF}$ value of SNe Ia and the nUV absolute magnitudes of
their host galaxies (Fig.~\ref{rmu}). Again, the distributions of
nUV absolute magnitudes of the host galaxies of the SNe Ia with
strong and weak CaII IR3 HVF are indistinguishable, i.e. a K-S
test shows that the sub-samples with $R_{\rm HVF}>0.2$ and $R_{\rm
HVF}<0.2$ have a probability of 46\% to be from the same mother
sample. However, it seems that most of SNe Ia locate in a
declining zonal, as dotted lines show. The results in this section
seem to be inconsistent with that discovered in
section~\ref{sect:3.5}, which could be derived from the fact that
the global parameters of a host galaxy is not a good tracer of
stellar population at the site of supernova explosion. The global
parameters of a host galaxy only represent the average information
of the stellar population in the host galaxy, and could erase the
intrinsic relation between SN Ia properties and their stellar
populations.

\begin{figure}
\centerline{\includegraphics[angle=270,scale=.38]{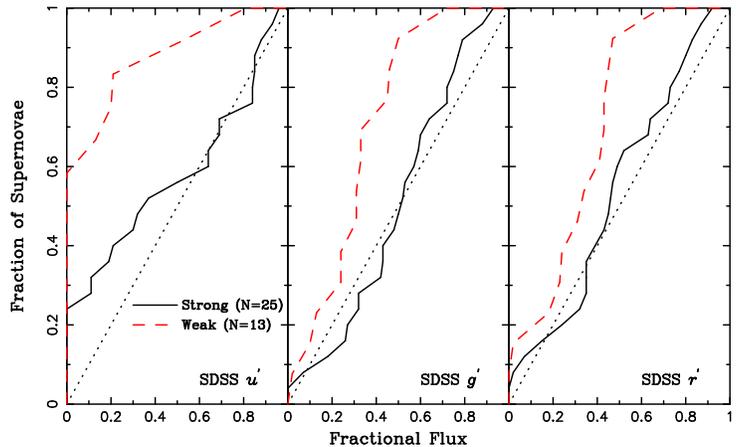}}
\caption{The cumulative distribution of the fractional flux of the
host galaxies at the SN explosion site for SNe Ia with strong
(black solid line) and weak (red dashed line) CaII IR3 HVF around
the maximum light, respectively, in SDSS $u^{\rm '}$ , $g^{\rm '}$
and $r^{\rm '}$ bands. The fractional flux data are from
\citet{WANGXF13}.}\label{fdis}
\end{figure}

\subsection{Pixel statistics}\label{sect:3.7}
The most direct method to determine the progenitor nature of an SN
Ia is to investigate its pre-explosion image (\citealt{LIWD11};
\citealt{MCCULLY14}). Such cases are rare and therefore the
statistics remain low, since it is only possible for events in
very nearby host galaxies. Another way is to investigate how the
properties of SNe Ia vary with different global parameters of
their host galaxies, as discussed in section~\ref{sect:3.6}.
However, this method could erase the intrinsic relation between SN
Ia properties and the information of their stellar populations,
because the global parameters of a host galaxy may only represent
the average information of the stellar population in the host
galaxy, which could be the reason why no correlation is found
between the strength of HVF and the global parameters of the host
galaxies in section~\ref{sect:3.6} and in \citet{ZHAOXL15}. An
intermediate method to constrain the nature of SN Ia progenitors
is to investigate the environments at the position of an SN Ia in
its host galaxy, e.g. the statistical analysis of a fractional
flux or a normalized cumulative rank (NCR) pixel value function of
the host galaxies at the SN explosion site (\citealt{FRUCHTER06};
\citealt{ANDERSON08}). Generally, the core-collapse SNe
approximately linearly trace the star-formation region (light) in
their host galaxies, while SNe Ia do not
(\citealt{ANDERSON15,ANDERSON15b}). Here, I will also use this
method to check whether or not there is a difference of the
environments between the SNe Ia with strong and weak CaII IR3 HVF
around the maximum light.

In Fig.~\ref{fdis}, I show the cumulative distribution of the
fractional flux of the host galaxies at SN explosion site for the
SNe Ia with strong and weak CaII IR3 HVF around the maximum light,
respectively, in SDSS $u^{\rm '}$ , $g^{\rm '}$ and $r^{\rm '}$
bands, where the fractional flux of an SN Ia represents the
fraction of total host light in pixels fainter than or equal to
the light in the pixel of SN Ia site in its host-galaxy image
(\citealt{FRUCHTER06}). A young population will trace the diagonal
line in the plot, while an old population will be far away from
the diagonal line. The figure shows that no matter what the SDSS
band, the cumulative distribution of the SNe Ia with strong CaII
IR3 HVF is closer to the line tracing the star-formation region
than that of the SNe Ia with weak HVF, although the distributions
for $g^{\rm '}$ and $r^{\rm '}$ are similar. A 2D K-S test for
$u^{\rm '}$ and $g^{\rm '}$ bands or $u^{\rm '}$ and $r^{\rm '}$
bands shows that the probability that the two sub-samples are from
the same mother sample is lower than 1.5\%, i.e. the SNe Ia with
strong maximum-light CaII IR3 HVF are more likely to be from the
relatively young stellar population, while those with weak
maximum-light CaII IR3 HVF tend to be from the relatively old
stellar populations. Whatever, the cumulative distributions for
$g^{\rm '}$ and $r^{\rm '}$ bands are much closer to the diagonal
line than those for $u^{\rm '}$ band, especially for the SNe Ia
with weak CaII IR3 HVF, which indicate that $g^{\rm '}$ and
$r^{\rm '}$ band lights would not be good tracers for the star
formation in a galaxy. However, nUV and ${\rm H\alpha}$ bands are
good choices (\citealt{ANDERSON15,ANDERSON15b}).

\begin{figure}
\centerline{\includegraphics[angle=270,scale=.38]{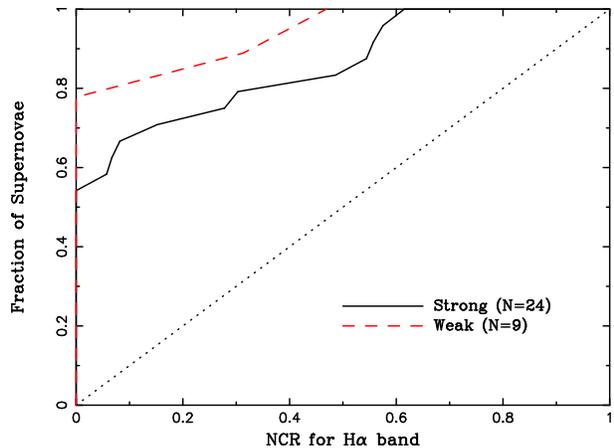}}
\caption{The cumulative NCR distributions of the SNe Ia with
strong (black solid line) and weak (red dashed line) CaII IR3 HVF
around the maximum light, respectively, for ${\rm H\alpha}$ band.
The NCR data of the host galaxies at SN Ia explosion site are from
\citet{ANDERSON15b}.}\label{ncrhalpha}
\end{figure}

\begin{figure}
\centerline{\includegraphics[angle=270,scale=.38]{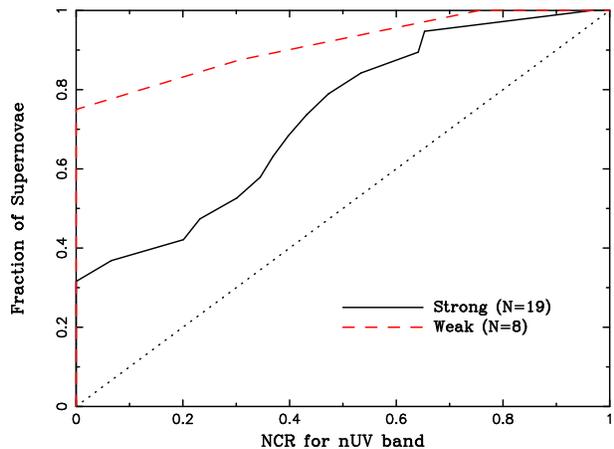}}
\caption{The cumulative NCR distributions of the SNe Ia with
strong (black solid line) and weak (red dashed line) CaII IR3 HVF
around maximum light, respectively, for nUV band. The NCR data of
the host galaxies at SN Ia explosion site are from
\citet{ANDERSON15b}.}\label{ncrnuv}
\end{figure}

The definition of an NCR value is different from the fractional
flux. The NCR value of a pixel is equal to the flux-count ratio
between this pixel and the one with the highest flux count within
the image of a host galaxy. Then, an NCR value is between 0 and 1,
where a value of 0 means that a pixel in the image is consistent
with zero flux or sky values, while a value of 1 means that the
pixel has the highest flux count in the image. Whether or not the
distribution of the NCR value for a given band in a cumulative
plot follows a diagonal one-to-one relation may provide
constraints on the population and progenitor properties of SNe Ia,
i.e. a very young population will trace the diagonal one-to-one
relation, while an old population will be far away from the
diagonal one-to-one relation. In another words, the closer the
cumulative line to the diagonal one-to-one relation, the younger
the population of SNe Ia (\citealt{ANDERSON08,ANDERSON15b}).
Figs.~\ref{ncrhalpha} and \ref{ncrnuv} present the cumulative NCR
distributions of the SNe Ia with strong and weak CaII IR3 HVF
around the maximum light for ${\rm H\alpha}$ and nUV bands,
respectively. These figures show that whatever the maximum-light
CaII IR3 HVF is strong or weak, or whatever the band is, the
cumulative distributions of the NCRs do not trace the diagonal
line, which indicates that the progenitors of SNe Ia do not belong
to very young population (see also
\citealt{ANDERSON15,ANDERSON15b}). This also verifies that the
SDSS $g^{\rm '}$ and $r^{\rm '}$ bands are not good tracers for
star formation. However, the cumulative distribution for the SNe
Ia with strong maximum-light CaII IR3 HVF is always closer to the
diagonal line than the one for the SNe Ia with weak maximum-light
CaII IR3 HVF, for both ${\rm H\alpha}$ and nUV bands, which
indicates that the populations of the SNe Ia with strong
maximum-light CaII IR3 HVF are relatively younger than those with
weak maximum-light CaII IR3 HVF. A 2D K-S test for the
distributions of NCR value for ${\rm H\alpha}$ and nUV bands
between the sub-samples of the SNe Ia with strong and weak CaII
IR3 HVF shows that the probability that the two sub-samples are
from the same mother sample is only 2.3\%, i.e. they arise from
different mother sample.

\begin{figure}
\centerline{\includegraphics[angle=270,scale=.38]{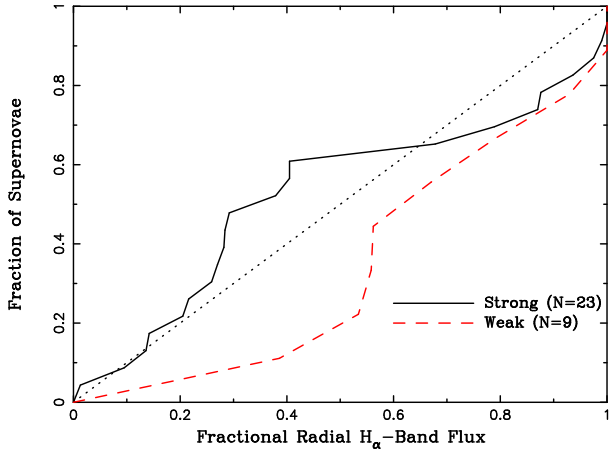}}
\caption{The cumulative distribution of Fr value for the SNe Ia
with strong (black solid line) and weak (red dashed line) CaII IR3
HVF around maximum light, respectively. The Fr data of the host
galaxies at SN Ia site are from
\citet{ANDERSON15b}.}\label{frhalpha}
\end{figure}

\subsection{Radial analysis}\label{sect:3.8}
The delay time\footnote{The delay time is the elapsed timescale
from the primordial system formation to the supernova explosion.
of an SNe Ia shares the similar meaning of its stellar population
to a great extent. An SN Ia with a long delay time belongs to old
population, and vice versa.} Comparing with the core-collapse SNe,
SNe Ia have a significant delay time from star formation to
explosion, and then they are very possible to explode at a
position far away from their birth sites. Since a stellar
population with different age and metallicity in a galaxy locates
at different characteristic galactocentric radial positions, I may
further investigate the environments of SNe Ia by exploring the
position of SNe Ia with respect to the radial distribution of
different stellar populations, e.g. a `Fr' fractional flux value
may provide such an information. Here, only ${\rm H\alpha}$ band
is considered (\citealt{ANDERSON15b}). A value of ${\rm Fr}=0$ for
an SN Ia means that the SN Ia locates at the central peak pixel in
the ${\rm H\alpha}$ band image of its host galaxy, while a value
of ${\rm Fr}=1$ implies that the SN Ia exploded at an outer region
of its host galaxy, where the ${\rm H\alpha}$ band flux is even
equal to the sky value (see details in \citealt{ANDERSON15b}).
Generally, in the cumulative plot for ${\rm Fr}$, the cumulative
fraction for a young population increases more quickly at low
${\rm Fr}$ value than an old population. In Fig.~\ref{frhalpha},
we present the cumulative distribution of the Fr value for the SNe
Ia with strong and weak CaII IR3 HVF around the maximum
brightness, respectively. As said above, the cumulative value for
the SNe Ia with strong maximum-light CaII IR3 HVF increases more
quickly at low ${\rm Fr}$ value than those with weak maximum-light
CaII IR3 HVF, i.e. the distributions indicate that the SNe Ia with
strong maximum-light CaII IR3 HVF belong to relatively younger
population than those with weak maximum-light CaII IR3 HVF. A K-S
test for the distributions of Fr value shows that the probability
that the two sub-samples are from the same mother sample is only
3.7\%, i.e. they are from different mother samples.

Here, the sample size of the SN Ia with weak maximum-light CaII
IR3 HVF is much smaller than that with strong maximum-light CaII
IR3 HVF, which is probably from the fact that all the host
galaxies in \citet{ANDERSON15b} are star-forming galaxies.
Actually, combining the small sample of the SN Ia with weak
maximum-light CaII IR3 HVF, this fact is another piece of evidence
to support our discovery, i.e. the SNe Ia with strong
maximum-light CaII IR3 HVF are more likely to correlate with a
relatively young population.

\begin{figure}
\centerline{\includegraphics[angle=270,scale=.38]{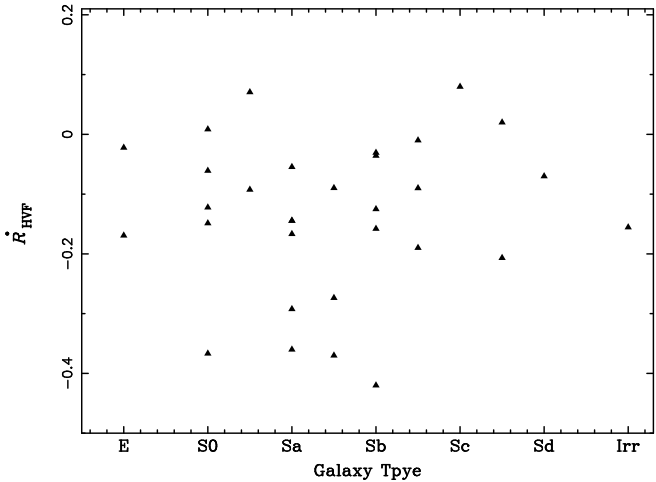}}
\caption{The decay rate of $R_{\rm HVF}$ for CaII IR3 absorption
line around maximum brightness versus host galaxy morphology,
where the data are from \citet{SILVERMAN12b} and
\citet{SILVERMAN15}.}\label{rdotgt}
\end{figure}

\subsection{Decay rate of $R_{\rm HVF}$}\label{sect:3.9}
Above sections have shown that the SNe Ia with very strong CaII
IR3 HVF around maximum light tend to occur in later type host
galaxies, and some statistical analysis also show that these SNe
Ia correlate with relatively younger stellar population than those
with weak CaII IR3 HVF. Since I only check the dependence of the
$R_{\rm HVF}$ values of SNe Ia at maximum brightness on their
exploding environments, our results might imply that SNe Ia in
relatively young environments have a slower decay rate of $R_{\rm
HVF}$ than those in old environments. By calculating the average
value of the various measurable parameters of SNe Ia,
\citet{ZHAOXL15} found that the decay rate of the HVFs roughly
correlates with $\Delta m_{\rm 15}(B)$ and $V_{\rm max}^{\rm Si}$.
So, since it is widely known that the bright SNe Ia favor late
type galaxy (\citealt{HAMUY96}; \citealt{WANGLF97}), we would
expect that the decay rate of the strength of the CaII IR3 HVF
correlates with the host galaxy morphology.

In Fig.~\ref{rdotgt}, I show the decay rate of $R_{\rm HVF}$ for
CaII IR3 absorption line ($\dot{R}_{\rm HVF}$) around the maximum
brightness as a function of the host galaxy morphology. Here, I do
not find any potential correlation between $\dot{R}_{\rm HVF}$ and
the host galaxy morphology. The difference from that in
\citet{ZHAOXL15} could be from that their decay rate is measured
at 7 days before maximum brightness, while measured around the
maximum brightness here. However, the statistical error bars of
the average values in \citet{ZHAOXL15} are so big that it could
also conclude that the decay rate of $R_{\rm HVF}$ does not depend
on $\Delta m_{\rm 15}(B)$ and $V_{\rm max}^{\rm Si}$, which is
consistent with my result here.

\begin{figure}
\centerline{\includegraphics[angle=270,scale=.38]{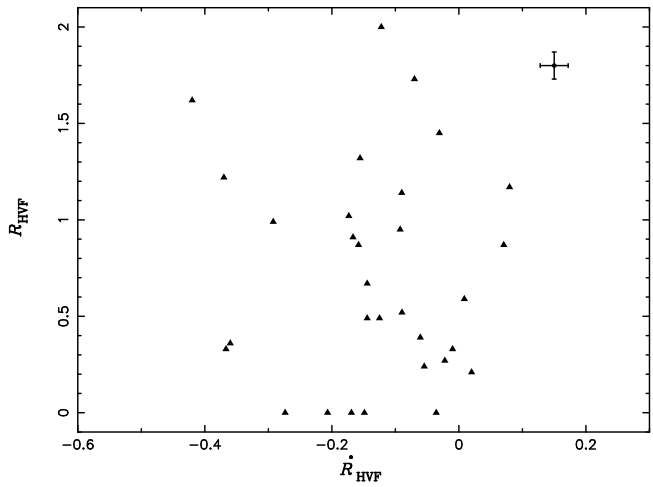}}
\caption{$R_{\rm HVF}$ versus the decay rate of $R_{\rm HVF}$ for
CaII IR3 absorption line around maximum brightness, where the data
are from \citet{SILVERMAN15}. The cross shows the typical error of
the data.}\label{rhdot}
\end{figure}

To confirm above conclusion, I still need to justify that
$\dot{R}_{\rm HVF}$ is independent in $R_{\rm HVF}$. We present
the plot of $R_{\rm HVF}$ versus the decay rate of $R_{\rm HVF}$
for CaII IR3 absorption line around maximum brightness in
Fig.~\ref{rhdot}. The figure shows that there is not any
potentially correlation between $\dot{R}_{\rm HVF}$ and $R_{\rm
HVF}$, as I expected. In addition, I do not find any potential
correlation between $\dot{R}_{\rm HVF}$ of CaII IR3 absorption
line and the REW of SiII 635.5 nm absorption line around maximum
brightness, either. For simplicity, I do not show the plot again
here. So there is indeed a correlation between the strength of
CaII IR3 HVF around maximum light in SNe Ia and their stellar
population.

In Figs.~\ref{rdotgt} and ~\ref{rhdot}, I notice that some SNe Ia
have a maximum-light $\dot{R}_{\rm HVF}$ value larger than 0, i.e.
the strength of CaII IR3 HVF becomes stronger after maximum. For
example, $R_{\rm HVF}=0.62\pm0.07$ at $t=-3.9$ day and $R_{\rm
HVF}=1.17\pm0.02$ at $t=3.0$ day for SN 1999dq, and $R_{\rm
HVF}=0.39\pm0.05$ at $t=-4.5$ day and $R_{\rm HVF}=0.87\pm0.10$ at
$t=2.3$ day for SN 2006bt (see table A4 in \citealt{SILVERMAN15}).
In general, the HVFs appear strongest in early-time spectra and
become weak with time, but some well observed SNe Ia show
interesting HVF evolutions. As one of the best observed SNe Ia,
i.e. SN 2011fe, the strength of the CaII IR3 HVF statistically
significantly increases with time during the first week after
supernova explosion, and then decreases with time after it reaches
to a maximum value (\citealt{CHILDRESS14}). The evolution of the
$R_{\rm HVF}$ of the CaII IR3 line in SN 2006X also shows a
similar behavior to SN 2011fe (see Table A4 in
\citealt{SILVERMAN15}). In addition, some SNe Ia show a plateau in
the evolution curve of $R_{\rm HVF}$, e.g. SN 2004dt, SN 2009ig
and SN 2012fr (\citealt{CHILDRESS14}; \citealt{ZHAOXL15}),
Especially, SN 2005cf presents both above behaviors
(\citealt{ZHAOXL15}). These behaviors on the evolution of the
strength of CaII IR3 HVF provide another constraint on the origin
of HVFs.

\section{DISCUSSION}\label{sect:4}
\subsection{Origins of the HVFs}\label{sect:4.1}
In this paper, based on the published data of CaII IR3 HVF around
maximum light in literatures, I found that the SNe Ia with strong
CaII IR3 HVF tend to occur in later type host galaxies or early
type galaxies with significant recent star formation, and some
statistical analysis also show these SNe Ia correlates with
relatively younger stellar population than those with weak CaII
IR3 HVF. So, my discoveries provide very strong constraints on the
origins of HVFs, even on the progenitor and explosion models of
SNe Ia. Generally, there are three popular scenarios to explain
the origins of the HVFs shown in the spectra of SNe Ia, i.e. the
abundance enhancement (AE), density enhancement (DE) and
ionization effect (IE) scenarios (\citealt{MAZZALI05a};
\citealt{TANAKA08}; \citealt{BLONDIN13}). Combining with previous
results in literatures, I will discuss which scenario is more
possible to explain my discoveries in the following sections.

\subsubsection{AE scenario}\label{sect:4.1.1}
For AE scenario, the outer layers of supernova ejecta are
dominated by Si and Ca, which implies that the outer layers of the
progenitor WD are significantly burned, or the significant burned
materials in inner region are brought up to the outer part of
supernova ejecta during the explosion phase
(\citealt{MAZZALI05a}).

Several mechanisms could contribute to above abundance structure.
For the delayed-detonation model (\citealt{KHOKHLOV91}), the first
deflagration phase may lead to asymmetry for the distribution of
detonation ignition point, which could burn C/O into Si/Ca in some
directions (\citealt{BLONDIN13}, \citealt{SEITENZAHL13}). A
detonation in the violent merger of two WDs or a gravitational
confined detonation on the surface of WD could also produce a
similar situation, i.e. a detonation in one side may produce more
high-velocity Si/Ca than the other side (\citealt{PLEWA04};
\citealt{KASEN05}; \citealt{PAKMOR12}). Theoretically, the
convection or a high accretion rate before the explosion may
naturally lead to an off-center ignition in a WD, and then an
asymmetric explosion (\citealt{KUHLEN06}; \citealt{CHENXC14}). It
has been suggested that the evolution of the photospheric velocity
may reflect the fact of an asymmetric explosion, i.e. LVG SNe Ia
are observed along the off-center-ignition side, while HVG SNe Ia
are observed from another side (\citealt{MAEDA10}). Polarization
observations show that supernova ejecta is not spherically
symmetric (\citealt{WANGLF08}). Especially it is found that the
higher the REW value of SiII 635.5 nm absorption line, the larger
its polarization, which may be explained by the delayed-detonation
mode (\citealt{MENGXC17}). However, it is still unclear how the
asymmetric explosion affects the HVFs in SNe Ia. In particular, we
did not find a difference of the $R_{\rm HVF}$ distribution
between LVG SNe Ia and HVG SNe Ia (see Fig.~\ref{disvg}), which
indicates that the asymmetric explosion could not affect the HVFs
in an SN Ia.

If the HVFs in SNe Ia are from above mechanisms, our discoveries
indicate that the progenitor population of an SN Ia would affect
its explosion, and require that the explosion from a relative
young population is more likely to produce Si/Ca at the outer
layers of supernova ejecta, and the layers have a higher velocity
(see Figs.~\ref{rvsi} and ~\ref{rva}), or a relatively young
population leads to a more asymmetric explosion. At present, it is
still unclear how the progenitor population of an SN Ia affect its
final explosion, although in principle the initial condition of an
exploding CO WD would be determined by its progeniotr. Maybe, the
cooling time of the initial WD before the onset of accretion or
the initial WD mass in a single-degenerate system might play a
kind of role (\citealt{CHENXC14}; \citealt{MENGXC18}).

An alternative possibility is from the double-detonation model, in
which a He detonation initiated near the WD surface may also
produce AE in this region and then show HVFs in the spectra of SNe
Ia (\citealt{WOOSLEY11}; \citealt{SHEN14}; \citealt{MAGUIRE14}).
This mechanism may naturally explain the fact that the
line-forming regions for HVF and PVF are detached. The companion
of the exploding CO WD may be a helium star or a helium WD.
Generally, the systems with helium star companions belong to
relative young population, and the systems with helium WDs tend to
be old population (\citealt{WANGB13}; \citealt{MENGXC15b}). So, if
the abundance enhanced Si/Ca from the double-detonation model is
the origin of the HVFs in SNe Ia, our discoveries imply that the
systems with helium star companions would produce the more
energetic SNe Ia with strong HVFs, while the systems with helium
WDs would lead to those with weak HVFs. However, present
double-detonation model can not give such a result and no evidence
shows that the detonations from helium stars and helium WDs are
different. Whatever, this is worth being carefully investigated in
the future (\citealt{TANIKAWA19}).

Another recently explored possibility is that a helium shell flash
on a white dwarf with mass close to the Chandrasekhar mass limit
could build up a layer enhanced in silicon or calcium pre-SN
explosion, where the helium shell is from the accretion of
hydrogen or helium from a non-degenerate companion in a
single-degenerate system (\citealt{KATO18}). The Si/Ca rich
material should locate in the most outer layers of supernova
ejecta, which would physically separate from the photospheric
silicon layers. This might explain the correlation between $R_{\rm
HVF}$ and $(\bar{v}_{\rm CI}-\bar{v}_{\rm Si})$, but can not
explain the SNe Ia with $R_{\rm HVF}=0$ and a high value of
$(\bar{v}_{\rm CI}-\bar{v}_{\rm Si})$. However, it is unclear how
this mechanism would relate to the stellar population of SNe Ia.
Might there be a difference on the helium flash between young and
old populations?

In summary, although the AE scenario may provide a natural way to
produce the HVFs, it is completely unclear how this scenario
correlates with the stellar population of SNe Ia. In particular,
the strength of the HVFs from the AE scenario should decrease with
time since the supernova ejecta become thinner and thinner with
time, which is quite difficult to explain the increase and plateau
behaviors in the $R_{\rm HVF}$ evolution curve with time presented
in some well observed SNe Ia, e.g. SN 2011fe, SN 2005cf and SN
2006X (\citealt{CHILDRESS14}; \citealt{ZHAOXL15};
\citealt{SILVERMAN15}). So, the origin of the HVFs in SNe Ia is
more possible to be from a mechanism relating to the progenitor
models of SNe Ia, rather than explosion models.

\subsubsection{DE and IE scenarios}\label{sect:4.1.2}
In the DE scenario, the HVFs originate from a high-density or
density bump shell (equivalent to adding mass) at outer layers, in
which the abundance is a typical value in the expanding ejecta
(\citealt{MAZZALI05a}; \citealt{TANAKA08};
\citealt{MULLIGAN17,MULLIGAN18}). Generally, two mechanisms are
suggested to produce the density variation. One is from an
asymmetrical explosion, as discussed in the above section. Then,
the ejecta in some directions may be affected by burning
differently from other directions. As discussed in
Sect.~\ref{sect:4.1.1}, such a mechanism is difficult to explain
the correlation between the strength of the maximum-light CaII IR3
HVF and the stellar population of SNe Ia, and is difficult to
explain the increase and plateau behavior of the strength
evolution of the CaII IR3 HVF shown in some SNe Ia. Another
mechanism to achieve density bump shell may be from the
interaction between supernova ejecta and relatively dense CSM
around progenitor system (\citealt{MULLIGAN17,MULLIGAN18}).

In the IE scenario, a little amount of hydrogen is mixed at the
outermost layer as a source of free electrons. Because Ca is
mostly doubly ionized at the outermost layer, the increased free
electron density suppress the ionization status of Ca by
recombination, and then the fraction of CaII is increased
(\citealt{MAZZALI05a}; \citealt{TANAKA08}). The hydrogen may be
from a contamination of hydrogen on the WD surface in a
single-degenerate system, or from the interaction between
supernova ejecta and relatively dense CSM as in DE scenario.

Both DE and IE scenarios have a potential ability, or play a role
together, to explain the correlation between the strength of HVFs
and the stellar populations of SNe Ia, if the HVFs are from the
interaction between the supernova ejecta and the CSM. The
interaction between supernova ejecta and the CSM may contribute to
the CaII IR3 HVF by the following three ways: I) adding mass at
the highest velocity region of supernova ejecta by the
accumulation of CSM; II)changing the ionization degree of calcium
by increasing the free electron density from hydrogen; III)
increasing the residence time of photons in the hydrogen shell by
a higher scattering-off probability of photons from a higher
electron density. All these effects contribute to an increased
line absorption at the highest-velocity region of supernova
ejecta, and then a HVF is presented in the absorption line of CaII
IR3 line in the early optical spectrum of an SN Ia. Because the
CaII line is the most remarkable feature in the optical spectrum
of an SN Ia, it is also most significantly affected by the
interaction between supernova ejecta and the CSM
(\citealt{MAZZALI05a}; \citealt{TANAKA08}). Generally, the young
SNe Ia tend to have a dense CSM, while the environments around old
SNe Ia are relatively clear, e.g. 1991T-like SNe Ia just belong to
relatively young population, while no CSM is discovered around old
1991bg-like SNe Ia (\citealt{JOHANSSON13b}; \citealt{FISHER15}).
Especially, these two scenarios have a potential ability to
explain the evolution of the strength of the HVFs with time. Here,
the dominant factor is the decreasing density of supernova ejecta
with time, which leads to the global decrease of the strength of
the HVFs with time. However, the CSM structure around SNe Ia may
be various, depending on the mass-loss histories of their
progenitors, e.g. a wind or a shell structure as some SNe Ia shown
(\citealt{PATAT07}; \citealt{DILDAY12}), which could result in
different evolution behaviors of the strength of the HVFs.

Whatever, two special kinds of SNe Ia must be paid more attention,
if the HVFs are from the interaction between the supernova ejecta
and the hydrogen-rich material, e.g. 2002cx-like and SN Ia-CSM
objects (\citealt{FOLEY13}; \citealt{SILVERMAN13}). Both kinds of
objects belong to relatively young population and have a much
denser CSM than normal SNe Ia (\citealt{CHOMIUK16};
\citealt{LYMAN18}; \citealt{SZALAI19}). \citet{MENGXC18b} even
suggested that these two peculiar subclasses of SNe Ia share the
same origin, i.e. from the hybrid CONe WD + MS system, based on a
new developed version of singe-degenerate model [common-envelope
wind (CEW) model, \citealt{MENGXC17a}]. However, the CaII IR3
absorption features in the spectra of these peculiar SNe Ia are
quite different from the normal SNe Ia. For SNe Ia-CSM, the CaII
IR3 features are strong and broad emission features, rather than
absorption lines in normal SNe Ia (\citealt{SILVERMAN13}). Such
difference between SNe Ia-CSM and the normal SNe Ia could arise
from the amount of hydrogen-rich material. To form the HVFs in
normal SNe Ia, a few $10^{\rm -3}$ ${\rm M}_{\odot}$ hydrogen-rich
materials are enough, but the amount of the CSM around SNe Ia-CSM
is much more massive than this value. For example, the amount of
the CSM around SN 2002ic (the prototype of SNe Ia-CSM,
\citealt{HAMUY03}) may be as massive as 0.5-6 ${\rm M}_{\odot}$
(\citealt{CHUGAI04}; \citealt{WANGLF04}; \citealt{KOTAK04}). For
2002cx-like SNe Ia, no significant HVF is discovered, but a
significantly lower maximum-light expansion velocity than normal
SNe Ia becomes a typical character (\citealt{FOLEY13}). The low
expansion velocity could result in an overlap or mix line-forming
region between the HVFs and PVFs, as shown in some normal SNe Ia
with high pEW of photospheric component, but without the
high-velocity CaII IR3 lines (see the discussions in
Sect.\,\ref{sect:3.3}). However, no definitive evidence supports
or denies above ideas at present, although SN 2006X provides a
clue for the ideas (see Sect.\,\ref{sect:3.3}). Especially, it is
still unclear how a large amount of hydrogen-rich amount affect
the CaII IR3 lines. Then, more efforts on these subjects are
needed in the future.

If the HVFs in SNe Ia arise from the interaction between supernova
ejecta and hydrogen-rich CSM and the CSM is massive enough, the
SNe Ia with very strong HVFs would show a variable sodium
absorption lines in their early spectra, as SN 2006X showed
(\citealt{PATAT07}; \citealt{SIMON09}; \citealt{BLONDIN09};
\citealt{FERRETTI16}). Recently, \citet{WANGXF19} provided a
sample of SNe Ia with variable sodium absorption line, including
SN 2006X, and these SNe Ia indeed present a strong CaII IR3 HVF.
For example, SN 2006X, the prototype of SNe Ia with a variable
sodium line, has a maximum-light value of $R_{\rm
HVF}=1.14\pm0.07$ [also for SN 1999dq, 2002bo and 2002cd, see
Table A4 in \citet{SILVERMAN15} and Table 2 in \citet{WANGXF19}].
In addition, SN 2002dj presents a quickly variable sodium line
around 8 days before maximum light (Fig. 2 in \citealt{WANGXF19}).
Interestingly, SN 2002dj also shows a very strong CaII IR3 HVF at
the same epoch, i.e. $R_{\rm HVF}=1.17\pm0.06$ (Table A4 in
\citealt{SILVERMAN15}). The CSM around SNe Ia may present
themselves by another way, i.e. the CSM may redden the color of
SNe Ia and then lead to a color excess, e.g. $E(B-V)$\footnote{The
CSM leading to the HVFs in SNe Ia is relatively close to the
exploding WD, while the CSM that is far away from the exploding WD
may also contribute to the color excess of SNe Ia.}. Recently,
\citet{BULLA18} found that some SNe Ia show time-variable $E(B-V)$
and these SNe Ia prefer the SNe Ia with a variable sodium line,
e.g. SN 2002bo, 2002cd and 2006X, which also show a very strong
maximum-light CaII IR3 HVF. Another interesting fact is that the
distribution of the $R_{\rm HVF}$ of the maximum-light CaII IR3
HVF is quite similar to the distribution of the $E(B-V)$ of SNe Ia
around maximum brightness, i.e. a low value peak of $R_{\rm HVF}$
[$E(B-V)$] with a long tail (\citealt{REINDL05};
\citealt{MENGXC09}).

\subsection{Progenitor of SNe Ia}\label{sect:4.2}
The correlation between the strength of the maximum-light CaII IR3
HVF and the stellar population of SNe Ia may provide meaningful
constraints on the progenitor models of SNe Ia. In particular,
combined with the results in literatures, our discoveries seems to
favor that the HVFs in SNe Ia arise from the interaction between
supernova ejecta and the CSM around SNe Ia. This is also helpful
to distinguish between different progenitor models of SNe Ia.
Generally, the progenitors of SNe Ia are categorized into three
main scenarios based on the companion nature of the exploding CO
WDs and on the explosion mechanism, i.e. the single-degenerate
(SD, \citealt{WI73}; \citealt{NTY84}), double-degenerate (DD,
\citealt{IT84}; \citealt{WEB84}), and sub-Chandrasekhar
double-detonation (D-DET) models (\citealt{WOOSLEY94};
\citealt{LIVNE95}; \citealt{SHEN13}). In the following sections,
\textbf{I} will discuss which model is more likely to explain all
the observations of the HVFs in SNe Ia.

\subsubsection{SD model}\label{sect:4.2.1}
In the SD model, the companion of the exploding CO WD may be a
main sequence (MS), a subgiant (SG), a red-giant (RG) or a helium
star. The accreted hydrogen-rich or helium-rich material is burned
on the surface of the CO WD into carbon and oxygen, and then
deposited onto the WD. When the WD mass reaches to a value colse
to the Chandrasekhar mass limit, an SN Ia may be produced. After
the supernova explosion, the companion may survive
(\citealt{WANGB12}; \citealt{MAOZ14}; \citealt{MENGXC15}). If the
correlation between the strength of the HVFs and the sellar
population arises from the interaction between supernova ejecta
and the hydrogen-rich CSM, the SD model has a potential ability to
explain the origin of the HVFs, where the CSM is from the outflow
or wind from the binary system. For a given initial WD, the amount
and density of the CSM is then mainly determined by the initial
companion mass, i.e. the more massive the companion, the more
likely to form a dense CSM by a wind (e.g. the common-envelope
wind, \citealt{MENGXC17a}). In addition, the age of an SN Ia from
the SD channel is also dominated by the companion mass, i.e. the
more massive the companion, the younger the SN Ia. So, a
relatively younger SN Ia means a relatively denser CSM, and then a
stronger HVFs, i.e. in principle, the SD model may explain my
discoveries in this paper. Since the CSM structure may be various
depending on the detailed mass-loss histories, the SD model may
also explain the increase or plateau behaviors in the CaII IR3
$R_{\rm HVF}$ evolution with time in principle. In addition, based
on the SD model, in principle, the CSM may exist anywhere from the
vicinity of the progenitor system of an SN Ia to a distance of
more than 300 pc to the system, depending on the detailed way that
the CSM is formed (\citealt{MENGXC17a,MENGXC18b}), as deduced from
observations (\citealt{BROERSEN14}; \citealt{BULLA18}).

A symbiotic system with a low-mass RG could also be the progenitor
of an SN Ia (\citealt{LI97}; \citealt{HAC99b};
\citealt{CHENXC11}). Such a system belongs to an old population,
but may also have a relatively dense CSM, which seems to be
inconsistent with our discovery. Maybe, a spin-up/spin-down
mechanism is necessary (\citealt{JUSTHAM11};
\citealt{DISTEFANO12}). For this mechanism, the CO WD will not
explode as an SN Ia immediately for a rapid rotation, spined up by
the accretion, even if its mass reaches to or exceeds the
Chandrasekhar mass limit. The rapid rotating WD must experience a
spin-down phase to explode, although the spin-down timescale is
quite uncertain (\citealt{DISTEFANO11}; \citealt{MENGXC13}).
During the spin-down phase, the environment around the progenitor
system may become very clean and the RG companion may become a dim
helium WD or sdB star (\citealt{JUSTHAM11}; \citealt{MENGXC19}).
Maybe, the initial WD mass could play a key role for the spin-down
timescale (\citealt{MENGXC18}). If so, the SD model is not
inconsistent with our discoveries.

The interacting CSM may emit at radio band and X-ray, but no any
SN Ia has been detected in radio or in X-ray bands, even for the
two most well observed SNe Ia, e.g. SN 2011fe and SN 2014J
(\citealt{MARGUTTI12,MARGUTTI14}; \citealt{CHOMIUK12};
\citealt{PEREZTORRES14}). These negative results indicate a very
low CSM density around the exploding SNe Ia, which is consistent
with the required amount of the hydrogen-rich CSM to explain the
HVF of SNe Ia, but without hydrogen emission lines in the spectra
of SNe Ia (e.g. a few times of $10^{\rm -3}$~$M_{\odot}$,
\citealt{MAZZALI05a}; \citealt{TANAKA08}). Whatever, if the amount
of hydrogen-rich CSM is massive enough, the hydrogen emission line
would expected, as shown in the spectra of SNe Ia-CSM
(\citealt{HAMUY03}; \citealt{DILDAY12}; \citealt{SILVERMAN13}).

\subsubsection{DD model}\label{sect:4.2.2}
In the DD model, a binary system consisting of two CO WDs loses
its orbital angular momentum by gravitational radiation, and
merges finally. If the total mass of the binary system exceeds the
Chandrasekhar mass limit, the merger may explode as an SN Ia.
Whatever, the merger will be disrupted completely and no surviving
companion exists after supernova explosion (\citealt{WANGB12};
\citealt{MAOZ14}; \citealt{MENGXC15}).

Actually, the CSM may also be formed from a DD system, but it is
hydrogen- and helium-deficient, i.e. it mainly consists of
carbon/oxygen, where the CSM could arise from a super wind during
the merging process (\citealt{SOKER13}). For example, although
they are very rare, the so-called `super-Chandrasekhar' SNe Ia are
suggested to be that a Chandrasekhar-mass WD explodes inside a
dense carbon/oxygen envelope which is the leftover of the WD-WD
merger (\citealt{HOWELL06}; \citealt{SCALZO14};
\citealt{TAUBENBERGER13,TAUBENBERGER19}). If such a kind of CSM
plays a role to form the HVFs in SNe Ia, the origin of HVFs would
be the DE scenario.

It was suggested that the younger, more massive stars produce more
massive white dwarfs, and then more massive carbon/oxygen envelope
to be formed during the merging process (\citealt{HOWEL11};
\citealt{MAOZ12}), which could contribute to the correlation
between the HVFs of SNe Ia and their stellar population. However,
the detailed binary population synthesis shows that the
distribution of the total masses of the DD systems is rather
uniform within the whole delay-time interval because the delay
time of SNe Ia from the DD systems is mainly determined by
gravitational wave radiation, rather than their progenitor
evolutionary time (\citealt{MENGYANG12}), which indicates that the
leftover of the WD-WD merger would be the same, whatever the
progenitors of SNe Ia belongs to young or old stellar population.
Maybe, as needed in the SD model, a spin-down timescale and a
magnetic field would be necessary, i.e. for young SNe Ia, the
magnetic field of the merger is stronger, and then the spin-down
timescale is shorter to form a relatively denser carbon/oxygen CSM
(\citealt{ILKOV12}). However, it is completely unclear why younger
DD systems would form a merger with stronger magnetic field and
then experience a shorter spin-down timescale.

The leftover of the WD-WD merger could have other geometric
structure, e.g. the disk-originated CSM, depending on the merger
timescale (\citealt{LEVANON15,LEVANON19}). The less massive WD is
tidally destroyed by its more massive companion and form an
accretion disk around the more massive companion. A bipolar wind
or jet might be blown off to form the disk-originated CSM, which
is also hydrogen- and helium-deficient. If the interaction between
supernova ejecta and the disk-originated CSM is the reason leading
to the HVFs in SNe Ia, the DE scenario would be the origin of the
HVFs. However, similar to above discussions, many efforts are
needed to explore why young SNe Ia have such disk-originated CSM,
while old SNe Ia do not, although the distribution of the total
mass of the DD systems leading to SNe Ia is rather uniform across
the whole delay-time interval (\citealt{MENGYANG12}).

Maybe, both the SD and DD models could produce SNe Ia together,
but with different age population, e.g. the SD model mainly
produces relatively young SNe Ia and the DD model mainly produces
old SNe Ia, since some evidence showed that different SNe Ia may
be from different population (\citealt{WANGXF13}). For such a
combination of different progenitor models, the IE scenario would
be the origin of the HVFs in SNe Ia to explain the correlation
between the strength of HVFs and the stellar population of SNe Ia.
However, the combination scenario is difficult to explain why
almost all SNe Ia show the HVF in very early phase, if the IE
scenario is the origin of the HVFs.

As a special case of DD model, the core-degenerate (CD) model may
also have an ability to explain the correlation between the
maximum-light CaII IR3 HVF and the stellar population of SNe Ia,
where the hydrogen-rich CSM is from the common envelope formed by
merger between a CO WD and an AGB star with an CO core.
(\citealt{KASHI11}). Whatever, similar to the SD model, a
spin-down timescale is also necessary to explain the correlation
found in this paper, where a magneto-dipole radiation torque
dominates the spin-down timescale (\citealt{ILKOV12}). However, it
is completely unclear why young merger tends to have a strong
magnetic field, and then lead to a short spin-down timescale
(\citealt{ILKOV12}). In addition, the CD model is less possible to
contribute to the majority of SNe Ia (\citealt{MENGYANG12};
\citealt{WANGB17}), although the argument on the contribution
exists (\citealt{ILKOV13}).

\subsubsection{D-DET model}\label{sect:4.2.3}
The D-DET model is also frequently discussed, in which the
companion of the CO WD is a helium WD or a helium star. The
companion fills its Roche lobe and a stable mass transfer occurs.
If the mass-transfer rate is not high enough so that the
helium-rich material can not be stably burning, the helium will
gradually accumulate on the CO WD. When the helium layer is thick
enough, a detonation will be ignited at the bottom of the helium
shell, where the inward supersonic detonation wave may lead to the
second detonation in the center of the CO WD. After the supernova
explosion, a hypervelocity companion may survive
(\citealt{GEIER15}; \citealt{SHEN18}).

The D-DET model could contribute to the HVFs by three different
ways. I) there is a thin hydrogen-rich layer on a helium WD, which
will transfer onto the WD prior to the He core's tidal disruption.
Finally, this hydrogen-rich material will be ejected from the
binary system via a way similar to a classical nova to form the
CSM (\citealt{SHEN13}). The interaction between supernova ejecta
and the CSM could contribute to the HVFs in SNe Ia. A key problem
for this scenario is whether the amount of the CSM is enough or
not, since the hydrogen-rich layer on a He WD is generally only
$2\times10^{\rm -4}-2\times10^{\rm -3}$ $M_{\odot}$ and a part of
hydrogen-rich material will be consumed during classical nova
explosion phase, while $4-5\times10^{\rm -3}$ $M_{\odot}$ hydrogen
rich material is necessary for IE scenario (\citealt{SHEN13};
\citealt{MAZZALI05a}; \citealt{TANAKA08}). Maybe, both DE an IE
scenarios play a role together to form the HVFs in SNe Ia here.
II) as discussed in Sect.~\ref{sect:4.1.1}, the first detonation
may directly produce high-velocity Si/Ca in the outer layer of
supernova ejecta (\citealt{SHEN14}). III) a dense helium shell may
exist prior to the complete explosion of the white dwarf, which is
ejected by the detonation on the surface of the white dwarf or
arises from the accretion on to the progenitor white dwarf
(\citealt{SHEN10}; \citealt{SHEN14}). The HVFs, especially the
CaII IR3 feature, may be explained by the interaction between
supernova ejecta and the dense helium shell
(\citealt{MULLIGAN17,MULLIGAN18}). Whatever the way is, it is
unclear why young SNe Ia produce strong HVFs, while old SNe Ia
produce weak HVFs for the D-DET model. The D-DET model is also
relatively difficult to explain the various evolution behaviors of
the $R_{\rm HVF}$ of CaII IR3 absorption line with time.

\subsection{Origin of the relation between $R_{\rm HVF}$ and $(\bar{v}_{\rm CI}-\bar{v}_{\rm Si})$}\label{sect:4.3}
In section~\ref{sect:3.3}, I found that there is very good linear
relation between $R_{\rm HVF}$ and $(\bar{v}_{\rm CI}-\bar{v}_{\rm
Si})$, and suggest $(\bar{v}_{\rm CI}-\bar{v}_{\rm Si})$ as an
indicator to judge whether or not there is a high-velocity
component in CaII IR3 absorption line. This relation could provide
important clues on explosion mechanism of SNe Ia and the origin of
the HVFs in SNe Ia.

From Figs.~\ref{rvsi} and \ref{rva}, CaII IR3 line usually has a
larger absorption-weighted velocity than SiII 635.5 nm line.
Considering that different elements locate different layers in
supernova ejecta (\citealt{FILIPPENKO97}, \citealt{MAZZALI07}),
these results indicate that calcium layer has a higher average
velocity than silicon layer. As discussed in above sections, some
mechanisms, e.g. asymmetric explosion and double detonation, could
contribute to the results, but seem to be difficult to explain the
correlation between the strength of CaII IR3 HVF and the
population of SNe Ia. In other words, these mechanisms have more
or less difficulties to simultaneously explain why $R_{\rm HVF}$
linearly increases with $(\bar{v}_{\rm CI}-\bar{v}_{\rm Si})$ and
why $R_{\rm HVF}$ depends on the stellar population of SNe Ia.

However, there might be a very simple explanation on the relation
in Fig.~\ref{rvsi}, i.e. the silicon and calcium layers have a
similar velocity distribution in the velocity space of supernova
ejecta, but the IE is the origin of CaII IR3 HVF. Because calcium
has  much lower number density than silicon in the forming region
of HVF and the calcium ion is more likely to be affected by free
election from hydrogen than silicon, a small amount of hydrogen
could lead to a more significant HVF in CaII IR3 line than in SiII
635.5 nm line, i.e. the weighted mean absorption velocity of CaII
IR3 line becomes larger than that of SiII 635.5 nm line (see the
discussion in \citealt{MAZZALI05a}). At the same time, the
strength of CaII IR3 HVF becomes larger. This simple explanation
is also relatively easy to explain why CaII IR3 absorption line
usually shows more significant HVF than SiII 635.5 nm line.
Whatever, more detailed efforts are needed to verify such an idea.

\section{CONCLUSIONS}\label{sect:5}
Although many data on the HVFs of SNe Ia have been published, the
origin of the HVFs is still unclear. Especially, no definitive
constraint on the progenitor or explosion model was obtained from
the HVF observation. In this paper, based on the data published in
literatures, I found strong evidence that the strength of the CaII
IR3 HVF around maximum brightness correlates with the stellar
population of SNe Ia, and the main results are as follows.

1) SNe Ia with strong CaII IR3 HVF around maximum brightness tend
to occur in late-type galaxies. Some lenticular galaxies also host
the SNe Ia with strong CaII IR3 HVF, but all of these lenticular
galaxies show the signature of recent or on-going star formation,
and the SNe Ia with strong maximum-light CaII IR3 HVF locate at or
close to the star formation regions in their host galaxies.

2) In the sample of \citet{SILVERMAN15}, all of the eight
1991T-like SNe Ia have a very strong CaII IR3 HVF around maximum
light, while among seventeen 1991bg-like SNe Ia, sixteen present
weak or no CaII IR3 HVF around maximum brightness. It is well
established that 1991T-like SNe Ia arise from young stellar
population and 1991bg-like SNe Ia belong to old stellar population

3) By pixel statistics, I found that the SNe Ia with strong CaII
IR3 HVF around maximum brightness show a higher degree of
association with star formation index, e.g. ${\rm H}\alpha$ or
near-UV emission emission, than those with weak maximum-light CaII
IR3 HVF.

4) Because all the host galaxies are star-forming galaxies in the
sample of \citet{ANDERSON15b}, the sample size of SNe Ia with weak
maximum-light CaII IR3 HVF is much smaller than that with strong
maximum-light CaII IR3 HVF (see Sect.\,\ref{sect:3.8}).

5) The distribution of the REW of SiII 635.5 nm around maximum
brightness between the SNe Ia with strong and weak CaII IR3 HVF
are quite different, i.e. SNe Ia with strong maximum-light CaII
IR3 HVF favor small or larger REW, while those with weak
maximum-light CaII IR3 HVF tend to have a middle value of REW.

6) \textbf{I} found that the strength of the CaII IR3 HVF around
maximum brightness is linearly dependent on the difference of the
absorption-weighted velocities between CaII IR3 and SiII 635.5 nm
absorption lines at the same phase. Then, I suggest that the
difference of the absorption-weighted velocities may be helpful to
judge whether or not there is a high-velocity component in the
CaII IR3 feature in the spectra of SNe Ia.

Based on the above results, I may get a conclusion that the SNe Ia
with strong CaII IR3 HVF around maximum brightness favor
relatively younger stellar population than those with weak
maximum-light CaII IR3 HVF, which provides meaningful constraints
on the explosion and progenitor models. Although a
spin-up/spin-down mechanism is necessary, the SD model may
relatively naturally explain the correlation between the strength
of the HVFs and the stellar population of SNe Ia discovered here,
while more efforts for the DD and D-DET models are necessary.
Although our results can not give a definitive conclusion on the
origin of the HVFs, the discoveries here seem to disfavor the AE
scenario as the origin of the HVFs, and the HVFs are less possible
to arise from the explosion itself of an SN Ia. On the contrary,
the interaction between supernova ejecta and the hydrogen-rich CSM
could be a natural origin of the HVFs in SNe Ia, where the DE and
IE scenarios could paly a role together.

\section*{Acknowledgments}
I am grateful to the anonymous referee for his/her constructive
comments which help me to improve the manuscript. I thank Zhengwei
Liu, Callum McCutcheon and Philipp Podsiadlowski for their helpful
discussions. This work was partly supported by the NSFC (Nos.
11973080 and 11733008), Yunnan Foundation (No. 2015HB096 and
2017HC018), and CAS (No. KJZD-EW-M06-01). X.M. thanks the support
by Yunnan Ten Thousand Talents Plan - Young \& Elite Talents
Project.


\begin{thebibliography}{}
\bibitem[\protect\citeauthoryear{Abbott et al.}{2019}]{ABBOTT19}
Abbott, T.M.C., Allam, S., Andersen, P. et al., 2019, ApJL, 872,
L30
\bibitem[\protect\citeauthoryear{Anderson \& James}{2008}]{ANDERSON08}
Anderson, J.P. \& James, P.A., 2008, MNRAS, 390, 1527
\bibitem[\protect\citeauthoryear{Anderson et al.}{2015a}]{ANDERSON15}
Anderson, J. P., James, P. A., Habergham, S. M., Galbany, L.,
Kuncarayakti, H., 2015a, PASA, 32, 19
\bibitem[\protect\citeauthoryear{Anderson et al.}{2015b}]{ANDERSON15b}
Anderson, J. P.£¬ James, P. A.£¬ F\"{o}rster, F. et al., 2015b,
MNRAS, 448, 732
\bibitem[\protect\citeauthoryear{Ann et al.}{2015b}]{ANN15}
Ann, H.B., Seo, M., Ha, D.K, 2015, ApJS, 217, 27
\bibitem[\protect\citeauthoryear{Balzano}{1983}]{BALZANO83}
Balzano, V.A., 1983, ApJ, 268, 602
\bibitem[\protect\citeauthoryear{Benetti et al.}{2005}]{BENETTI05}
Benetti, S., Cappellaro, E., Mazzali, P. A. et al., 2005, ApJ,
623, 1011
\bibitem[\protect\citeauthoryear{Blondin et al.}{2009}]{BLONDIN09}
Blondin, S., Prieto, J.L., Patat, F. et al., 2009, ApJ, 693, 207
\bibitem[\protect\citeauthoryear{Blondin et al.}{2013}]{BLONDIN13}
Blondin S., Dessart L., Hillier D. J., Khokhlov A. M., 2013,
MNRAS, 429, 2127
\bibitem[\protect\citeauthoryear{Bulla et al.}{2018}]{BULLA18}
Bulla, M., Goobar, A., Dhawan, S., 2018, MNRAS, 479, 3663
\bibitem[\protect\citeauthoryear{Branch et al.}{2009}]{BRANCH09}
Branch, D., Dang, L. C., \& Baron, E. 2009, PASP, 121, 238
\bibitem[\protect\citeauthoryear{Broersen et al.}{2014}]{BROERSEN14}
Broersen, S., Chiotellis, A., Vink, J., Bamba, A., 2014, MNRAS,
441, 3040
\bibitem[\protect\citeauthoryear{Cameron}{2011}]{CAMERON11}
Cameron, E., 2011, PASA, 28, 128
\bibitem[\protect\citeauthoryear{Chugai \& Yungelson}{2004}]{CHUGAI04}
Chugai, N.N., Yungelson L.R., 2004, Astronomy Letters, 30, 65
\bibitem[\protect\citeauthoryear{Chen et al.}{2011}]{CHENXC11}
Chen X., Han, Z., Tout, C. A., 2011, ApJL, 735, L31
\bibitem[\protect\citeauthoryear{Chen et al.}{2014}]{CHENXC14}
Chen X., Han, Z., Meng, X., 2014, MNRAS, 438, 3358
\bibitem[\protect\citeauthoryear{Chevalier}{1990}]{CHEVALIER}
Chevalier, R. A., 1990, in Supernovae, ed. A. G. Petschek (New
York: Springer-Verlag), 91
\bibitem[\protect\citeauthoryear{Childress et al.}{2014}]{CHILDRESS14}
Childress M. J., Filippenko A. V., Ganeshalingam M., Schmidt B.
P., 2014, MNRAS, 437, 338
\bibitem[\protect\citeauthoryear{Chomiuk et al.}{2012}]{CHOMIUK12}
Chomiuk, L., Soderberg, A., Moe, M., et al. 2012, ApJ, 750, 164
\bibitem[\protect\citeauthoryear{Chomiuk et al.}{2016}]{CHOMIUK16}
Chomiuk, L., Soderberg, A. M., Chevalier, R. A., et al. 2016, ApJ,
821, 119
\bibitem[\protect\citeauthoryear{Della Valle \& Livio}{1994}]{DELLA94}
Della Valle, M. \& Livio, M., 1994, ApJ, 423, L31
\bibitem[\protect\citeauthoryear{Dilday et al.}{2012}]{DILDAY12}
Dilday B., Howell D.A., Cenko S.B. et al., 2012, Science, 337, 942
\bibitem[\protect\citeauthoryear{Di Stefano et al.}{2011}]{DISTEFANO11}
Di Stefano R., Voss R., \& Claeys J.S.W., 2011, ApJL, 738, L1
\bibitem[\protect\citeauthoryear{Di Stefano \& Kilic}{2012}]{DISTEFANO12}
Di Stefano R., \& Kilic M. 2012, ApJ, 759, 56
\bibitem[\protect\citeauthoryear{Filippenko}{1997}]{FILIPPENKO97}
Filippenko, A.V., 1997, ARA\&A, 35, 309
\bibitem[\protect\citeauthoryear{Ferretti et al.}{2016}]{FERRETTI16}
Ferretti, R., Amanullah, R., Goobar, A. et al., 2016, A\&A, 592,
A40
\bibitem[\protect\citeauthoryear{Fisher \& Jumper}{2015}]{FISHER15}
Fisher, R. \& Jumper, K., 2015, ApJ, 805, 150
\bibitem[\protect\citeauthoryear{Foley et al.}{2013}]{FOLEY13}
Foley, R. J., Challis, P. J., Chornock, R., et al. 2013, ApJ, 765,
57
\bibitem[\protect\citeauthoryear{Fruchter et al.}{2006}]{FRUCHTER06}
Fruchter, A. S., Levan, A. J., Strolger, L. et al., 2006, Nature,
441, 463
\bibitem[\protect\citeauthoryear{Gall et al.}{2018}]{GALL18}
Gall, C., Stritzinger, M. D., Ashall, C. et al., 2018, A\&A, 611,
A58
\bibitem[\protect\citeauthoryear{Geier et al.}{2015}]{GEIER15}
Geier, S., F\"{u}rst, F., Ziegerer, E. et al., 2015, Science, 347,
1126
\bibitem[\protect\citeauthoryear{Goobar \& Leibundgut}{2011}]{GOOBAR11}
Goobar, A. \& Leibundgut, B., 2011, ARNPS, 61, 251
\bibitem[\protect\citeauthoryear{Hachisu et al.}{1999}]{HAC99b}
Hachisu, I., Kato, M., Nomoto, K., 1999, ApJ, 522, 487
\bibitem[\protect\citeauthoryear{Hamuy et al.}{1996}]{HAMUY96}
Hamuy M., Phillips M.M., Schommer R.A. et al., 1996, AJ, 112, 2391
\bibitem[\protect\citeauthoryear{Hamuy et al.}{2003}]{HAMUY03}
Hamuy, M.,  Phillips, M. M., Suntzeff, N. B. et al., 2003, Nature,
424, 651
\bibitem[\protect\citeauthoryear{Hatano et al.}{1999}]{HATANO99}
Hatano, K., Branch, D., Fisher, A., Baron, E., Filippenko, A.V.,
1999, ApJ, 525, 881
\bibitem[\protect\citeauthoryear{Hillebrandt \& Niemeyer}{2000}]{HN00}
Hillebrandt, W., Niemeyer, J.C., 2000, ARA\&A, 38, 191
\bibitem[\protect\citeauthoryear{Hillebrandt et al.}{2013}]{HILLEBRANDT13}
Hillebrandt, W., Kromer, M., R\"{o}pke, F. K., Ruiter, A. J.,
2013, FrPhy, 8, 116
\bibitem[\protect\citeauthoryear{Howell}{2001}]{HOWELL01}
Howell, D. A. 2001, ApJ, 554, L193
\bibitem[\protect\citeauthoryear{Howell et al.}{2006}]{HOWELL06}
Howell D.A., Sullivan, M., Nugent, P.E. et al., 2006, Nature, 443,
308
\bibitem[\protect\citeauthoryear{Howell}{2011}]{HOWEL11}
Howell, D.A., 2011, NatCo, 2E, 350
\bibitem[\protect\citeauthoryear{Hoyle \& Fowler}{1960}]{HF60}
Hoyle, F. \& Fowler, W.A., 1960, ApJ, 132, 565
\bibitem[\protect\citeauthoryear{Iben \& Tutukov}{1984}]{IT84}
Iben, I., Tutukov, A.V., 1984, ApJS, 54, 335
\bibitem[\protect\citeauthoryear{Ilkov \& Soker}{2012}]{ILKOV12}
Ilkov, M., Soker, N., 2012, MNRAS, 419, 1695
\bibitem[\protect\citeauthoryear{Ilkov \& Soker}{2013}]{ILKOV13}
Ilkov, M., Soker, N., 2013, MNRAS, 428, 579
\bibitem[\protect\citeauthoryear{Jha et al.}{2019}]{JHA19}
Jha, S.W., Maguire, K. \& Sullivan, M., 2019, NatAs, 3, 706
\bibitem[\protect\citeauthoryear{Johansson et al.}{2013a}]{JOHANSSON13}
Johansson, J., Thomas, D., Pforr, J. et al., 2013a, MNRAS, 435,
1680
\bibitem[\protect\citeauthoryear{Johansson et al.}{2013b}]{JOHANSSON13b}
Johansson, J., Amanullah, R., Goobar, A., 2013b, MNRAS, 431, L43
\bibitem[\protect\citeauthoryear{Justham}{2011}]{JUSTHAM11}
Justham S., 2011, ApJL, 730, L34
\bibitem[\protect\citeauthoryear{Kashi \& Soker}{2011}]{KASHI11}
Kashi, A. \& Soker N. 2011, MNRAS, 417, 1466
\bibitem[\protect\citeauthoryear{Kasen et al.}{2003}]{KASEN03}
Kasen D., Nugent, P., Wang, L. et al., 2003, ApJ, 593, 788
\bibitem[\protect\citeauthoryear{Kasen \& Plewa}{2005}]{KASEN05}
Kasen, D., Plewa, T., 2005, ApJ, 622, L41
\bibitem[\protect\citeauthoryear{Kato et al.}{2018}]{KATO18}
Kato, M., Saio, H., Hachisu, I., 2018, ApJ, 863, 125
\bibitem[\protect\citeauthoryear{Khokhlov}{1991}]{KHOKHLOV91}
Khokhlov, A.M., 1991, A\&A, 245, 114
\bibitem[\protect\citeauthoryear{Kotak et al.}{2004}]{KOTAK04}
Kotak, R., Meikle, W.P.S., Adamson, S. et al., 2004, MNRAS, 354,
L13
\bibitem[\protect\citeauthoryear{Kuhlen et al.}{2006}]{KUHLEN06}
Kuhlen, M., Woosley, S.E. \& Glatzmaier, G.A., 2006, ApJ, 640, 407
\bibitem[\protect\citeauthoryear{Levanon et al.}{2015}]{LEVANON15}
Levanon, N., Soker, N., \& Garc\'{i}a-Berro, E., 2015, MNRAS, 447,
2803
\bibitem[\protect\citeauthoryear{Levanon \& Soker}{2019}]{LEVANON19}
Levanon, N., Soker, N., 2019, ApJL, 872, L7
\bibitem[\protect\citeauthoryear{Li \& van den Heuvel}{1997}]{LI97}
Li, X.D., van den Heuvel, E.P.J., 1997, A\&A, 322, L9
\bibitem[\protect\citeauthoryear{Li et al.}{2011}]{LIWD11}
Li, W.-D., Bloom, J. S., Podsiadlowski, Ph. et al., 2011, Nature,
480, 348
\bibitem[\protect\citeauthoryear{Livne \& Arnett}{1995}]{LIVNE95}
Livne, E., \& Arnett, D., 1995, ApJ, 452, 62
\bibitem[\protect\citeauthoryear{Lyman et al.}{2018}]{LYMAN18}
Lyman, J. D., Taddia, F., Stritzinger, M. D., et al. 2018, MNRAS,
473, 1359
\bibitem[\protect\citeauthoryear{Mannucci et al.}{2005}]{MANNUCCI05}
Mannucci, F., Della Valle, M., Panagia, N. et al., 2005, A\&A,
433, 807
\bibitem[\protect\citeauthoryear{Maoz \& Mannucci}{2012}]{MAOZ12}
Maoz, D., \& Mannucci, F., 2012, PASA, 29, 447
\bibitem[\protect\citeauthoryear{Maoz, Mannucci \& Nelemans}{2014}]{MAOZ14}
Maoz D., Mannucci F., Nelemans G., 2014, ARA\&A, 52, 107
\bibitem[\protect\citeauthoryear{Maeda et al.}{2010}]{MAEDA10}
Maeda, K., Benetti, S., Stritzinger, M. et al., 2010, Nature, 466,
82
\bibitem[\protect\citeauthoryear{Maguire et al.}{2012}]{MAGUIRE12}
Maguire K., Sullivan, M., Ellis, R.S. et al., 2012, MNRAS, 426,
2359
\bibitem[\protect\citeauthoryear{Maguire et al.}{2014}]{MAGUIRE14}
Maguire K., Sullivan, M., Pan, Y.-C. et al., 2014, MNRAS, 444,
3258
\bibitem[\protect\citeauthoryear{Maguire et al.}{2018}]{MAGUIRE18}
Maguire, K., Sim, S.A., Shingles, L. et al., 2018, MNRAS, 477,
3567
\bibitem[\protect\citeauthoryear{Margutti et al.}{2012}]{MARGUTTI12}
Margutti, R., Soderberg, A. M., Chomiuk, L., et al. 2012, ApJ,
751, 134
\bibitem[\protect\citeauthoryear{Margutti et al.}{2014}]{MARGUTTI14}
Margutti, R., Parrent, J., Kamble, A., et al. 2014, ApJ, 790, 52
\bibitem[\protect\citeauthoryear{Marino et al.}{2011}]{MARINO11}
Marino, A., Rampazzo, R., Bianchi, L. et al., 2011, MNRAS, 411,
311
\bibitem[\protect\citeauthoryear{Marion et al.}{2013}]{MARION13}
Marion, G. H., Vinko, J., Wheeler, J. C. et al., 2013, ApJ, 777,
40
\bibitem[\protect\citeauthoryear{Maund et al.}{2013}]{MAUND13}
Maund, J. R., Spyromilio, J., H\"{o}flich, P. A.  et al., 2013,
MNRAS, 433, L20
\bibitem[\protect\citeauthoryear{Mazzali et al.}{2005a}]{MAZZALI05a}
Mazzali P.A., Benetti S., Stehle M. et al., 2005a, MNRAS, 357, 200
\bibitem[\protect\citeauthoryear{Mazzali et al.}{2005b}]{MAZZALI05b}
Mazzali P.A., Benetti S., Altavilla G. et al., 2005b, ApJL, 623,
L37
\bibitem[\protect\citeauthoryear{Mazzali et al.}{2007}]{MAZZALI07}
Mazzali, P,A., R\"{o}pke, F,K., Benetti, S, Hillebrandt, W, 2007,
Science, 315, 825
\bibitem[\protect\citeauthoryear{McCully et al.}{2014}]{MCCULLY14}
McCully, C., Jha, S. W., Foley, R. J. et al., 2014, Nature, 512,
54
\bibitem[\protect\citeauthoryear{Meng et al.}{2009}]{MENGXC09}
Meng, X., Chen, X., Han, Z., Yang, W., 2009, RA\&A, 9, 1259,
\bibitem[\protect\citeauthoryear{Meng \& Yang}{2012}]{MENGYANG12}
Meng X., \& Yang W., 2012, A\&A, 543, A137
\bibitem[\protect\citeauthoryear{Meng \& Podsiadlowski}{2013}]{MENGXC13}
Meng, X. \& Podsiadlowski, Ph., 2013, ApJL, 778, L35
\bibitem[\protect\citeauthoryear{Meng et al.}{2015}]{MENGXC15}
Meng X., Gao Y., Han Z., 2015, IJMPD, 24, 14, 1530029
\bibitem[\protect\citeauthoryear{Meng \& Han}{2015}]{MENGXC15b}
Meng, X., Han, Z., 2015, A\&A, 573, A57
\bibitem[\protect\citeauthoryear{Meng \& Podsiadlowski}{2017}]{MENGXC17a}
Meng, X. \& Podsiadlowski, Ph., 2017, MNRAS, 469, 4763
\bibitem[\protect\citeauthoryear{Meng et al.}{2017}]{MENGXC17}
Meng, X., Zhang, J., Han, Z., 2017, ApJ, 841, 62
\bibitem[\protect\citeauthoryear{Meng \& Han}{2018}]{MENGXC18}
Meng, X., Han, Z., 2018, ApJL, 855, L18
\bibitem[\protect\citeauthoryear{Meng \& Podsiadlowski}{2018}]{MENGXC18b}
Meng, X. \& Podsiadlowski, Ph., 2018, ApJ, 861, 127
\bibitem[\protect\citeauthoryear{Meng \& Li}{2019}]{MENGXC19}
Meng, X. \& Li, J., 2019, MNRAS, 482, 5651
\bibitem[\protect\citeauthoryear{Mulligan \& Wheeler}{2017}]{MULLIGAN17}
Mulligan, B.W., Wheeler, J.C., 2017, MNRAS, 467, 778
\bibitem[\protect\citeauthoryear{Mulligan \& Wheeler}{2018}]{MULLIGAN18}
Mulligan, B.W., Wheeler, J.C., 2018, MNRAS, 476, 1299
\bibitem[\protect\citeauthoryear{Navasardyan et al.}{2001}]{NAVASARDYAN01}
Navasardyan, H., Petrosian, A.R., Turatto, M., Cappellaro, E.,
Boulesteix, J., 2001, MNRAS, 328, 1181
\bibitem[\protect\citeauthoryear{Nomoto, Thielemann \& Yokoi}{1984}]{NTY84}
Nomoto, K., Thielemann, F-K., Yokoi, K., 1984, ApJ, 286, 644
\bibitem[\protect\citeauthoryear{Pakmor et al.}{2012}]{PAKMOR12}
Pakmor, R., Kromer, M., Taubenberger, S. et al., 2012, ApJ, 747,
L10
\bibitem[\protect\citeauthoryear{Pan et al.}{2015}]{PANYC15}
Pan, Y.-C., Sullivan, M., Maguire, K. et al., 2015, MNRAS, 446,
354
\bibitem[\protect\citeauthoryear{Parrent et al.}{2012}]{PARRENT12}
Parrent J., Howell, D. A., Friesen, B. et al., 2012, ApJ, 752, L26
\bibitem[\protect\citeauthoryear{Pastorello et al.}{2007}]{PASTROELLO07}
Pastorello, A., Taubenberger, S., Elias-Rosa, N. et al., 2007, MN,
376, 1301
\bibitem[\protect\citeauthoryear{Patat et al.}{2007}]{PATAT07}
Patat, E., Chandra, P., Chevalier, R., et al., 2007, Science, 317,
924
\bibitem[\protect\citeauthoryear{Patat et al.}{2009}]{PATAT09}
Patat F., Baade D., H\"{o}flich P. et al. 2009, A\&A, 508, 229
\bibitem[\protect\citeauthoryear{P\'{e}rez-Torres et al.}{2014}]{PEREZTORRES14}
P\'{e}rez-Torres, M., Lundqvist, P., Beswick, R., et al. 2014,
ApJ, 792, 38
\bibitem[\protect\citeauthoryear{Planck Collaboration et al.}{2018}]{PLANCK18}
Planck Collaboration, Aghanim, N., Akrami, Y., et al. 2018,
arXiv:1807.06209
\bibitem[\protect\citeauthoryear{Plewa et al.}{2004}]{PLEWA04}
Plewa, T., Calder, A. C., \& Lamb, D. Q. 2004, ApJL, 612, L37
\bibitem[\protect\citeauthoryear{Perlmutter et al.}{1997}]{PER97}
Perlmutter, S., Gabi, S., Goldhaber, G., et al., 1997, ApJ, 483,
565
\bibitem[\protect\citeauthoryear{Perlmutter et al.}{1999}]{PER99}
Perlmutter, S., Aldering, G., Goldhaber, G. et al., 1999, ApJ,
517, 565
\bibitem[\protect\citeauthoryear{Phillips}{1993}]{PHILLIPS93}
Phillips M.M., 1993, ApJ, 413, L105
\bibitem[\protect\citeauthoryear{Reindl}{2005}]{REINDL05}
Reindl, B., Tammann, G. A., Sandage, A. et al. 2005, ApJ, 624, 532
\bibitem[\protect\citeauthoryear{Riess et al.}{1996}]{RIESS96}
Riess, A., Press, W. H., Kirshner, R. P., 1996, ApJ, 473, 88
\bibitem[\protect\citeauthoryear{Riess et al.}{1998}]{RIE98}
Riess, A., Filippenko, A. V., Challis, P., et al., 1998, AJ, 116,
1009
\bibitem[\protect\citeauthoryear{Riess et al.}{2016}]{RIESS16}
Riess A.G., Macri L.M., Hoffmann S.L. et al., 2016, ApJ, 826, 56
\bibitem[\protect\citeauthoryear{Salim et al.}{2005}]{SALIM05}
Salim, S., Charlot, S., Rich, R.M. et al., 2005, ApJ, 619, L39
\bibitem[\protect\citeauthoryear{Scalzo et al.}{2014}]{SCALZO14}
Scalzo, R. A., Ruiter, A. J., Sim, S. A., 2014, MNRAS, 445, 2535
\bibitem[\protect\citeauthoryear{Schawinski et al.}{2007}]{SCHAWINSKI07}
Schawinski, K., Kaviraj, S., Khochfar, S. et al., 2007, ApJS, 173,
512
\bibitem[\protect\citeauthoryear{Seitenzahl et al.}{2013}]{SEITENZAHL13}
Seitenzahl, I.R., Ciaraldi-Schoolmann, F., R\"{o}pke, F.K. et al.
2013, MNRAS, 429, 1156
\bibitem[\protect\citeauthoryear{Shen et al.}{2010}]{SHEN10}
Shen, K.J., Kasen, D., Weinberg, N.N., Bildsten, L., Scannapieco,
E., 2010, ApJ, 715, 767
\bibitem[\protect\citeauthoryear{Shen et al.}{2013}]{SHEN13}
Shen, K.J., Guillochon, J., Foley, R.J., 2013, ApJ, 770, L35.
\bibitem[\protect\citeauthoryear{Shen \& Moore}{2014}]{SHEN14}
Shen, K.J., Moore, K., 2014, ApJ, 797, 46
\bibitem[\protect\citeauthoryear{Shen et al.}{2018}]{SHEN18}
Shen, K.J., Boubert, D.m, G\"{a}nsicke, B.T. et al., 2018, ApJ,
865, 15
\bibitem[\protect\citeauthoryear{Sheardown et al.}{2018}]{SHEARDOWN18}
Sheardown, A., Roediger, E., Su, Y. et al., 2018, ApJ, 865, 118
\bibitem[\protect\citeauthoryear{Silverman et al.}{2012a}]{SILVERMAN12}
Silverman, J. M., Kong, J. J., \& Filippenko, A. V. 2012a, MNRAS,
425, 1819
\bibitem[\protect\citeauthoryear{Silverman et al.}{2012b}]{SILVERMAN12b}
Silverman, J. M., Foley, R J., Filippenko, A.V. et al., 2012b,
MNRAS, 425, 1789
\bibitem[\protect\citeauthoryear{Silverman et al.}{2013}]{SILVERMAN13}
Silverman, J. M., Nugent, P. E., Gal-Yam, A. et al., 2013, ApJS,
207, 3
\bibitem[\protect\citeauthoryear{Silverman et al.}{2015}]{SILVERMAN15}
Silverman, J. M.; Vink\'{o}, J., Marion, G.H. et al., 2015, MNRAS,
451, 1973
\bibitem[\protect\citeauthoryear{Simon et al.}{2009}]{SIMON09}
Simon, J.D., Gal-Yam, A., Gnat, O., et al., 2009. ApJ 702, 1157
\bibitem[\protect\citeauthoryear{Smith et al.}{2010}]{SMITH10}
Smith, B.J., Giroux, M.L., Struck, C., Hancock, M., 2010, AJ, 139,
1212
\bibitem[\protect\citeauthoryear{Soker}{2013}]{SOKER13}
Soker, N., 2013, IAUS, 281, 72
\bibitem[\protect\citeauthoryear{Sternberg et al.}{2011}]{STERNBERG11}
Sternberg, A., Gal-Yam, A., Simon, J. D. et al., 2011, Science,
333, 856
\bibitem[\protect\citeauthoryear{Su et al.}{2011}]{SUYY17}
Su, Y., Kraft, R.P., Roediger, E. et al., 2017, ApJ, 834, 74
\bibitem[\protect\citeauthoryear{Sullivan et al.}{2006}]{SULLIVAN06}
Sullivan, M., Le Borgne, D., \& Pritchet, C. J., et al. 2006, ApJ,
648, 868
\bibitem[\protect\citeauthoryear{Sullivan et al.}{2011}]{SULLIVAN11}
Sullivan, M., Guy, J., Conley, A. et al., 2011, ApJ, 737, 102
\bibitem[\protect\citeauthoryear{Szalai et al.}{2019}]{SZALAI19}
Szalai, T., Zs\'{i}ros, S., Fox, O.D., Pejcha, O., M\"{u}ller, T.,
2019, ApJS, 241, 38
\bibitem[\protect\citeauthoryear{Tanaka et al.}{2008}]{TANAKA08}
Tanaka M., Mazzali, P.A., Benetti, S. et al., 2008, ApJ, 677, 448
\bibitem[\protect\citeauthoryear{Tanikawa et al.}{2019}]{TANIKAWA19}
Tanikawa, A., Nomoto, K., Nakasato, N., Maeda, K., 2019, ApJ, in
press, arXiv: 1909.09770
\bibitem[\protect\citeauthoryear{Taubenberger et al.}{2013}]{TAUBENBERGER13}
Taubenberger, S., Kromer, M., Hachinger, S. et al., 2013, MNRAS,
432, 3117
\bibitem[\protect\citeauthoryear{Taubenberger et al.}{2019}]{TAUBENBERGER19}
Taubenberger1, S., Floers1, A., Vogl, C. et al., 2019, in press,
arXiv: 1907, 6753
\bibitem[\protect\citeauthoryear{V\'{e}ron-Cetty \& V\'{e}ron}{2006}]{VERON06}
V\'{e}ron-Cetty, M.-P. \& V\'{e}ron, P., 2006, A\&A, 455, 773
\bibitem[\protect\citeauthoryear{Wang et al.}{1997}]{WANGLF97}
Wang, L., H\"{o}flich, P., Wheeler, J.C., 1997, ApJ, 483, L29
\bibitem[\protect\citeauthoryear{Wang et al.}{2003}]{WANGLF03}
Wang, L., Baade, D., H\"{o}flich, P. et al., 2003, ApJ, 591, 1110
\bibitem[\protect\citeauthoryear{Wang et al.}{2004}]{WANGLF04}
Wang L., Baade D., H\"{o}flich P. et al., 2004, ApJ, 604, L53
\bibitem[\protect\citeauthoryear{Wang et al.}{2006}]{WANGLF06}
Wang L., Baade D., H\"{o}flich P. et al. 2006, ApJ, 653, 490
\bibitem[\protect\citeauthoryear{Wang \& Wheeler}{2008}]{WANGLF08}
Wang, L., \& Wheeler, J. C. 2008, ARA\&A, 46, 433
\bibitem[\protect\citeauthoryear{Wang et al.}{2009}]{WANGXF09}
Wang, X., Filippenko, A. V., Ganeshalingam, M. et al., 2009, ApJ,
699, L139
\bibitem[\protect\citeauthoryear{Wang et al.}{2013}]{WANGXF13}
Wang, X., Wang, L., Filippenko, A. V., et al. 2013, Sci, 340, 170
\bibitem[\protect\citeauthoryear{Wang et al.}{2019}]{WANGXF19}
Wang, X., Chen, J., Wang, L. et al., 2019, ApJ, in press, arXiv:
1810.11936
\bibitem[\protect\citeauthoryear{Wang \& Han}{2012}]{WANGB12}
Wang, B., Han, Z., 2012, NewAR, 56, 122
\bibitem[\protect\citeauthoryear{Wang et al.}{2013}]{WANGB13}
Wang, B., Justham, S., Han, Z., 2013, A\&A, 559, A94
\bibitem[\protect\citeauthoryear{Wang et al.}{2017}]{WANGB17}
Wang, B., Zhou, W.-H., Zuo, Z.-Y. et al., 2017, MNRAS, 464, 3965
\bibitem[\protect\citeauthoryear{Webbink}{1984}]{WEB84}
Webbink, R.F., 1984, ApJ, 277, 355
\bibitem[\protect\citeauthoryear{Whelan \& Iben}{1973}]{WI73}
Whelan, J., Iben, I., 1973, ApJ, 186, 1007
\bibitem[\protect\citeauthoryear{Woosley \& Weaver}{1994}]{WOOSLEY94}
Woosley, S.E. \& Weaver, T.A., 1994, ApJ, 423, 371.
\bibitem[\protect\citeauthoryear{Woosley \& Kasen}{2011}]{WOOSLEY11}
Woosley, S.E., \& Kasen, D., 2011, ApJ, 734, 38
\bibitem[\protect\citeauthoryear{Zhang et al.}{2016}]{ZHANGJJ16}
Zhang, J., Wang, X., Sasdelli, M. et al., 2016, ApJ, 817, 114
\bibitem[\protect\citeauthoryear{Zhao et al.}{2015}]{ZHAOXL15}
Zhao, X., Wang, X., Maeda, K. et al., 2015, ApJS, 220, 20
\end{thebibliography}
\end{document}